\documentclass[sigconf, nonacm]{acmart}

\newcommand\vldbdoi{XX.XX/XXX.XX}
\newcommand\vldbpages{XXX-XXX}
\newcommand\vldbvolume{14}
\newcommand\vldbissue{1}
\newcommand\vldbyear{2020}
\newcommand\vldbauthors{\authors}
\newcommand\vldbtitle{\shorttitle} 
\newcommand\vldbavailabilityurl{URL_TO_YOUR_ARTIFACTS}
\newcommand\vldbpagestyle{plain} 

\usepackage{algpseudocodex}
\usepackage{algorithm}
\usepackage{graphicx}
\usepackage{textcomp}
\usepackage{xcolor}

\usepackage{amsthm}
\usepackage{booktabs}
\newtheorem{definition}{Definition}
\usepackage{cleveref}
\usepackage{multirow}
\usepackage[shortlabels,inline]{enumitem}
\usepackage[normalem]{ulem}
\usepackage{xspace}

\usepackage{xspace}

\usepackage{mathtools}

\newcommand{\SelectEntryPoint}{\textsc{SelectEntryPoint}\xspace}

\newcommand{\ClusBtreeOpen}{\textsc{Open}\xspace}
\newcommand{\ClusBtreeNext}{\textsc{Next}\xspace}

\newcommand{\GraphOpen}{\textsc{Open}\xspace}
\newcommand{\GraphNext}{\textsc{Next}\xspace}
\newcommand{\GraphNextFiltered}{\textsc{Next}\xspace}
\newcommand{\ExpandSearch}{\textsc{EnlargeSearch}\xspace}
\newcommand{\OneHopExpand}{\textsc{OneHopExpand}\xspace}
\newcommand{\TwoHopExpand}{\textsc{TwoHopExpand}\xspace}

\newcommand{\CompassSearch}{\textsc{CompassSearch}\xspace}

\newcommand{\Search}{\textsc{Search}\xspace}
\newcommand{\DistFunc}{\textsc{Dist}\xspace}

\newcommand{\MaintainQueues}{\textsc{Visit}\xspace}

\renewcommand{\Top}{\textsc{Top}\xspace}
\newcommand{\Push}{\textsc{Push}\xspace}
\newcommand{\Pop}{\textsc{Pop}\xspace}
\newcommand{\Size}{\textsc{Size}\xspace}

\newcommand{\NotEmpty}{\textsc{NotEmpty}\xspace}

\newcommand{\efc}{\textit{K}\xspace}

\newcommand{\deltaefs}{\textit{stepsize}\xspace}

\newcommand{\clusbtree}{\mathcal{B}\xspace}
\newcommand{\graph}{\mathcal{G}\xspace}
\newcommand{\cg}{\mathcal{G}'\xspace}
\newcommand{\indices}{\textit{Indices}\xspace}

\newcommand{\relq}{\textit{RelQ}\xspace}

\newcommand{\bg}{\textit{beg}\xspace}
\newcommand{\ed}{\textit{end}\xspace}

\newcommand{\passed}{\textit{passed}\xspace}

\newcommand{\rz}{\textit{TopQ}\xspace}
\newcommand{\cq}{\textit{CandiQ}\xspace}
\newcommand{\sq}{\textit{SharedQ}\xspace}
\newcommand{\recycleq}{\textit{RecycQ}\xspace}
\newcommand{\result}{\textit{ResQ}\xspace}
\newcommand{\va}{\textit{Visited}\xspace}
\newcommand{\efi}{\textit{efi}\xspace}
\newcommand{\ef}{\textit{ef}\xspace}
\newcommand{\efs}{\textit{efs}\xspace}

\newcommand{\cnt}{\textit{cnt}\xspace}

\newcommand{\currobj}{\textit{record}\xspace}
\newcommand{\currdist}{\textit{dist}\xspace}

\renewcommand{\break}{\textbf{break}}

\usepackage{color}
\usepackage{tcolorbox}
\definecolor{lightgray}{gray}{0.75}
\newtcolorbox{mybox}[3][]
{
  colframe = #2!25,
  colback  = #2!10,
  coltitle = #2!20!black,  
  title    = {#3},
  #1,
}

\usepackage{cleveref}

\makeatletter
\newcommand{\resplabel}[1]{\label[response]{resp:#1}}
\makeatother
\crefname{response}{Our Response}{Our Responses}

\newcounter{metarev}[meta]

\newcounter{reviewer}

\newcounter{revcomment}[reviewer]

\newcounter{reveval}[reviewer]

\newcounter{revrev}[reviewer]

\newcounter{response}[reviewer]

\renewcommand\theresponse{\thereviewer.\arabic{response}}

\theoremstyle{plain}

\theoremstyle{definition}
\newtheorem{example}{Example}

\newcommand{\name}{\textsc{Compass}}
\newcommand{\stitle}[1]{\vspace*{0.4em}\noindent{\bf #1.\/}}

\newcommand{\squishlist}{
	\begin{list}{$\bullet$}
		{ \setlength{\itemsep}{1pt}
			\setlength{\parsep}{1pt}
			\setlength{\topsep}{2.5pt}
			\setlength{\partopsep}{0.5pt}
			\setlength{\leftmargin}{1em}
			\setlength{\labelwidth}{1em}
			\setlength{\labelsep}{0.6em}
		}
	}
	\newcommand{\squishend}{
	\end{list}
}

\newcounter{msg}

\begin{document}
\title{Compass: General Filtered Search across Vector and Structured Data}

\author{Chunxiao Ye}
\affiliation{%
  \institution{The Chinese University of Hong Kong}
  \city{Hong Kong}
  \country{China}
}

\author{Xiao Yan}
\affiliation{%
  \institution{Wuhan University}
  \city{Wuhan}
  \country{China}
}

\author{Eric Lo}
\affiliation{%
  \institution{The Chinese University of Hong Kong}
  \city{Hong Kong}
  \country{China}
}

\begin{abstract}
The increasing prevalence of hybrid vector and relational data necessitates efficient, general support for queries that combine high-dimensional vector search with complex relational filtering. However, existing filtered search solutions are fundamentally limited by specialized indices, which restrict arbitrary filtering and hinder integration with general-purpose DBMSs.
This work introduces \textsc{Compass}, a unified framework that enables general filtered search across vector and structured data without relying on new index designs. Compass leverages established index structures -- such as HNSW and IVF for vector attributes, and B+-trees for relational attributes -- implementing a principled cooperative query execution strategy that coordinates candidate generation and predicate evaluation across modalities.
Uniquely, Compass maintains generality by allowing arbitrary conjunctions, disjunctions, and range predicates, while ensuring robustness even with highly selective or multi-attribute filters. Comprehensive empirical evaluations demonstrate that Compass consistently outperforms NaviX and ACORN, the few existing performant general frameworks, across diverse hybrid query workloads. It also matches the query throughput of specialized single-attribute indices in their favoring settings with only a single attribute involved, all while maintaining full generality and DBMS compatibility.
Overall, Compass offers a practical and robust solution for achieving truly general filtered search in vector database systems.

\end{abstract}

\maketitle

\pagestyle{\vldbpagestyle}
\begingroup\small\noindent\raggedright\textbf{PVLDB Reference Format:}\\
\vldbauthors. \vldbtitle. PVLDB, \vldbvolume(\vldbissue): \vldbpages, \vldbyear.\\
\href{https://doi.org/\vldbdoi}{doi:\vldbdoi}
\endgroup
\begingroup
\renewcommand\thefootnote{}\footnote{\noindent
This work is licensed under the Creative Commons BY-NC-ND 4.0 International License. Visit \url{https://creativecommons.org/licenses/by-nc-nd/4.0/} to view a copy of this license. For any use beyond those covered by this license, obtain permission by emailing \href{mailto:info@vldb.org}{info@vldb.org}. Copyright is held by the owner/author(s). Publication rights licensed to the VLDB Endowment. \\
\raggedright Proceedings of the VLDB Endowment, Vol. \vldbvolume, No. \vldbissue\ %
ISSN 2150-8097. \\
\href{https://doi.org/\vldbdoi}{doi:\vldbdoi} \\
}\addtocounter{footnote}{-1}\endgroup

\ifdefempty{\vldbavailabilityurl}{}{
\vspace{.3cm}
\begingroup\small\noindent\raggedright\textbf{PVLDB Artifact Availability:}\\
The source code, data, and/or other artifacts have been made available at \url{https://github.com/chunxy/Compass}.
\endgroup
}

\section{Introduction}
The rapid proliferation of unstructured data—including images, videos, and free-form documents—has precipitated the rise of \emph{vector databases}, wherein high-dimensional embeddings enable semantic retrieval via approximate $k$-nearest neighbor (A-$k$NN) search. These systems mark a significant shift in query capabilities, allowing similarity-based access beyond simple exact matches. However, practical workloads increasingly demand queries that jointly reason over semantic similarity and structured relational predicates: for example, retrieving “products similar to a reference item but priced below \$100,” or “images analogous to a query example but timestamped after 2020.” Addressing such requirements necessitates \emph{filtered search}, integrating vector-search and attribute-filtering in the same query.

Despite recent efforts, most existing filtered search solutions \cite{wangEfficientRobustFramework2023, wuHQANNEfficientRobust2022a, zuoSeRFSegmentGraph2024, engelsApproximateNearestNeighbor2024, mohoneyHighThroughputVectorSimilarity2023, xuIRangeGraphImprovisingRangededicated2024, liangUNIFYUnifiedIndex2025, jiangDIGRADynamicGraph2025, zhangEfficientDynamicIndexing2025, caiNavigatingLabelsVectors2024, pengDynamicRangeFilteringApproximate2025} remain ad-hoc and fragile under \emph{general filtering conditions}. The majority design specialized indices that tightly couple the vector embedding with one designated relational attribute, delivering high efficiency for specific, fixed filter types. Yet such approaches are fundamentally limited. They cannot support general relational filtering—encompassing numeric range predicates, multi-attribute queries, and complex conjunctions or disjunctions—unless ad-hoc pre- and post-filtering steps are introduced. 
As a result, their performance degrades when handling multiple attributes, or varied predicate combinations. Moreover, each index must pre-select the target relational attribute during index build time, resulting in one specialized index per relational attribute --- a solution that is neither scalable nor space-efficient.

Within published literature, ACORN~\cite{patelACORNPerformantPredicateAgnostic2024} and NaviX~\cite{sehgalNaviXNativeVector2025} are the notable exceptions, distinguished by their generality and seamless integration with database management systems. They decouple vector and relational indexing, upholding compatibility with general query processing. Nevertheless, they are hindered by a core limitation: relational filters disrupt the traversal connectivity of graph-based vector indices such as HNSW~\cite{malkovEfficientRobustApproximate2020}, as many neighbors are pruned by predicate evaluation. To compensate, they expand traversal to explore beyond the immediate neighbors, regaining coverage but paying the price in overhead and reduced query throughput (QPS).

This paper introduces \textsc{Compass}, a versatile filtered search framework that seamlessly integrates efficiency, robustness, and compatibility with DBMS. Rather than creating new specialized indices, Compass leverages established indices, such as HNSW and IVF for vector attributes, and B+-trees or even learned indices~\cite{gre, pgm} for relational attributes, which 
are all already battle-tested and adopted in industrial systems.
The key innovation of Compass is its \emph{shared candidate queue}, which facilitates cooperative query execution across these indices. The vector index operates mostly as usual, while the system dynamically supplements candidates from relational indices that meet the necessary filters when required. This architecture enables Compass to efficiently expand the search space while rigorously enforcing relational constraints, all without compromising generality or ease of integration.

Empirical results demonstrate that Compass consistently outperforms NaviX and ACORN across a wide range of query patterns, including single- and multi-attribute filters, varying selectivities, and both conjunctions and disjunctions. Remarkably, Compass achieves throughput comparable to that of specialized single-attribute indices even in scenarios that favor such indices, involving only one relational attribute. This is accomplished while maintaining full generality and leveraging proven database indexing components.

\begin{table*}[!t]
\centering
\caption{A summary of existing methods for filtered vector search and comparison with our \name.}
\label{table:feature}
\resizebox{\textwidth}{!}{
\begin{tabular}{@{}lcccccccccc@{}}
\toprule
\multicolumn{1}{c}{} & \multicolumn{4}{c}{Discrete Attribute Support}            & \multicolumn{3}{c}{Continuous Attribute Support} & \multicolumn{3}{c}{Index Structure Property}                               \\ \cmidrule(lr){2-5} \cmidrule(lr){6-8} \cmidrule(lr){9-11}
Method               & Equality     & Comparison   & Conjunction  & Disjunction  & Range          & Conjunction    & Disjunction    & Insertion    & Index Size                   & Build Time                   \\ \midrule
FilteredDiskANN      & $\checkmark$ & $\times$     & $\times$     & $\checkmark$ & $\times$       & $\times$       & $\times$       & $\checkmark$ & Moderate                     & Normal                       \\
DSG                  & $\checkmark$ & $\checkmark$ & $\square$    & $\square$    & $\checkmark$   & $\square$      & $\square$      & $\checkmark$ & Large                        & Long                         \\
iRangeGraph          & $\checkmark$ & $\checkmark$ & $\square$    & $\square$    & $\checkmark$   & $\square$      & $\square$      & $\times$     & Large                        & Long                         \\
SeRF                 & $\checkmark$ & $\checkmark$ & $\square$    & $\square$    & $\checkmark$   & $\square$      & $\square$      & $\times$     & Moderate                     & Long                         \\
{Weaviate} & $\checkmark$ & $\checkmark$ & $\checkmark$ & $\checkmark$ & $\checkmark$   & $\checkmark$   & $\checkmark$   & $\checkmark$ & Small & Short \\
{Milvus} & $\checkmark$ & $\checkmark$ & $\checkmark$ & $\checkmark$ & $\checkmark$   & $\checkmark$   & $\checkmark$   & $\checkmark$ & Small & Short \\
NaviX                & $\checkmark$ & $\checkmark$ & $\checkmark$ & $\checkmark$ & $\checkmark$   & $\checkmark$   & $\checkmark$   & $\checkmark$ & Normal                       & Short                       \\
ACORN                & $\checkmark$ & $\checkmark$ & $\checkmark$ & $\checkmark$ & $\checkmark$   & $\checkmark$   & $\checkmark$   & $\checkmark$ & Moderate & Moderate \\
\textbf{Compass}     & $\checkmark$ & $\checkmark$ & $\checkmark$ & $\checkmark$ & $\checkmark$   & $\checkmark$   & $\checkmark$   & $\checkmark$ & Normal                       & Normal                       \\ \bottomrule
\end{tabular}
}
\begin{flushleft} %
\small %
\centering
$\checkmark$: Full Support ~~~~~
$\square$: Partial Support ~~~~~
$\times$: No Support \\
\end{flushleft}

\end{table*}

\section{Preliminaries} \label{sec:preliminary}
\subsection{Problem Definitions}

In this work, we follow the mainstream setting adopted by prior filtered search works \cite{zuoSeRFSegmentGraph2024, xuIRangeGraphImprovisingRangededicated2024, patelACORNPerformantPredicateAgnostic2024, pengDynamicRangeFilteringApproximate2025} and focus on main-memory–resident datasets.

\begin{definition}
\textbf{General Filtered Search}. Let $
\mathcal{D} = \{ (v_i, a_i) \mid v_i \in \mathbb{R}^d,\, a_i \in \mathcal{A} \}
$
be a dataset where each record consists of a vector representation \( v_i \in \mathbb{R}^d \) and a tuple of relational attributes \( a_i \) defined over schema \( \mathcal{A} \).
A filtered query is defined as 
$
Q = (q, p),
$
where \( q \in \mathbb{R}^d \) is the query vector and \( p: \mathcal{A} \to \{\text{true}, \text{false}\} \) is a Boolean predicate over the attributes in \( \mathcal{A} \), composed of conjunctions and/or disjunctions of attribute conditions.
Let \( \delta: \mathbb{R}^d \times \mathbb{R}^d \to \mathbb{R} \) be a distance function.
Then, the general filtered search (GFS) problem is to find the set
\[
S^*_k = \operatorname*{Top-}k \bigl\{ (v_i, a_i) \in \mathcal{D}' \text{ ranked by } \delta(q, v_i) \bigr\}.
\]
where 
\[
\mathcal{D}' = \{ (v_i, a_i) \in \mathcal{D} \mid p(a_i) = \text{true} \}
\]
is the subset of records that satisfy the predicate \( p \).

\end{definition}

As in previous work, since exact vector search requires $O(n)$ time, we tackle the approximate GFS.

\begin{definition}
Approximate GFS returns a set $S_k \subseteq \{(v, a) \in \mathcal{D} \mid p(a) = \text{true}\}$, $|S_k| = k$, and the quality of the approximate result set is measured by recall $R = {|S_k \cap S_k^*|}/{|S_k^*|}$.
\end{definition}    

Our goal is to reach a high recall (e.g., 0.85 or 0.95) with a short processing time or a small number of vector distance computations.

\subsection{Indices for Vector Search}

\begin{figure}[!t]
    \centering
    \includegraphics[width=0.95\linewidth]{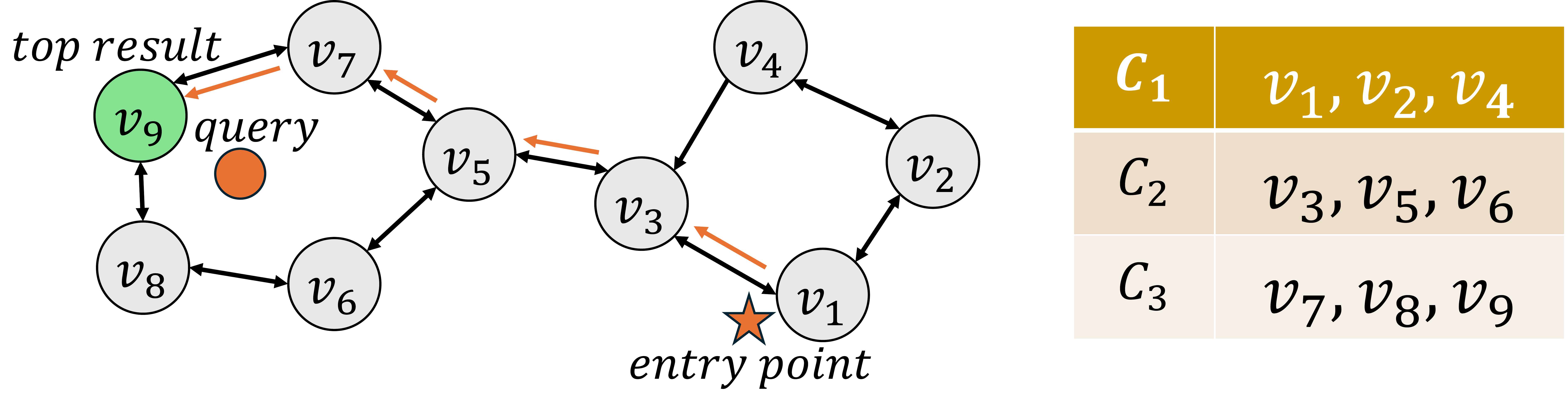}
    \caption{An illustration of proximity graph (left) and IVF index (right).}
    \label{fig:ivf-graph}
\end{figure}

Two types of indices are popular for vector search, i.e., proximity graph~\cite{malkovEfficientRobustApproximate2020, fuFastApproximateNearest2019, jayaramsubramanyaDiskANNFastAccurate2019, fuHighDimensionalSimilarity2022} and inverted file (IVF)~\cite{manningIntroductionInformationRetrieval2008, douzePolysemousCodes2016, babenkoInvertedMultiindex2012, jegouProductQuantizationNearest2011}. We provide an illustration of them in \Cref{fig:ivf-graph}.

In a proximity graph index, the nodes are the vectors and edges connect similar vectors. Vector search is conducted by a best-first-search-style graph traversal, which starts with a random or fixed entry node and uses a candidate queue to manage the node to visit and a top queue to manage nodes visited. 
When visiting a node, the graph traversal computes the distances between the query and all neighbors of the node and adds these neighbors to the candidate queue; then, an unvisited node with the smallest distance is selected from the candidate queue as the next to visit. Both search cost and result quality are controlled by the size of the top queue (usually denoted as $\textit{efs}$), with a larger queue size leading to more distance computations but higher result quality. There are many variants of the proximity graph index, e.g., HNSW~\cite{malkovEfficientRobustApproximate2020}, NSG~\cite{fuFastApproximateNearest2019}, Vamanna~\cite{jayaramsubramanyaDiskANNFastAccurate2019}, and SSG~\cite{fuHighDimensionalSimilarity2022}; they mainly differ in the entry point selection and edge pruning rule; while the graph traversal procedures for query processing are similar. 

The inverted file (IVF) index partitions a dataset's vectors into clusters (e.g. via K-means), with each cluster represented by a centroid. During query processing, all centroids are initially evaluated. Subsequently, only vectors within the top-ranking clusters (i.e., those with the centroids nearest to the query vector) are inspected as candidate results.
While proximity graphs generally exhibit higher search efficiency—requiring fewer distance computations for a given target recall—IVF offers the advantages of larger data access granularity and more regular data access patterns, rendering it highly amenable to parallelization.

\section{Related Work}
In this section, we give an overview of work related to general filtered search.
\Cref{table:feature} gives a summary and we elaborate them as follows.

\subsection{Specialized Indices for Label Filtering}
Early work on filtered search focused exclusively on the equality comparison of \emph{discrete} attributes, typically referred to as label filtering~\cite{wuHQANNEfficientRobust2022a, wangEfficientRobustFramework2023, gollapudiFilteredDiskANNGraphAlgorithms2023}. Among them, \cite{wangEfficientRobustFramework2023} and~\cite{wuHQANNEfficientRobust2022a} fuse the relational attributes into the vector to build the graph on the fused representation.
FilteredDiskANN \cite{gollapudiFilteredDiskANNGraphAlgorithms2023}
extends the graph-based ANN structure by applying a label-aware pruning strategy—which ensures path navigability for filtered queries.
At query time, FilteredDiskANN dynamically maintains a priority queue of candidates by iteratively adding only those neighboring nodes that satisfy the query's label filter predicate.

\subsection{Specialized Indices for 1D Range Filtering}
Recent advances~\cite{mohoneyHighThroughputVectorSimilarity2023, engelsApproximateNearestNeighbor2024, xuIRangeGraphImprovisingRangededicated2024, zuoSeRFSegmentGraph2024, pengDynamicRangeFilteringApproximate2025, zhangEfficientDynamicIndexing2025, jiangDIGRADynamicGraph2025, liangUNIFYUnifiedIndex2025} extend to support continuous attribute but limit the number of attributes to one.
Particularly, Super-Post-filtering~\cite{engelsApproximateNearestNeighbor2024} proposes partitioning the relational attribute domain using segment tree and building separate graph index for each segment.
iRangeGraph~\cite{xuIRangeGraphImprovisingRangededicated2024} streamlines such index construction by dynamically composing only the subgraphs relevant to a given query range. Both Super-Post-filtering and iRangeGraph, however, require maintaining separate graph structures for each segment. As attribute cardinality or range granularity increases, the storage overhead becomes substantial, making these methods impractical for scenarios involving large attribute domains or fine-grained filtering requirements. In our experiments, they typically resulted in an index size of up to $6\times$ the original index. Other partitioning methods either face the similar problem of bloated index size~\cite{zhangEfficientDynamicIndexing2025, jiangDIGRADynamicGraph2025, liangUNIFYUnifiedIndex2025} or rely on statically-maintained partitions subject to query workload change~\cite{mohoneyHighThroughputVectorSimilarity2023, engelsApproximateNearestNeighbor2024}.

SeRF~\cite{zuoSeRFSegmentGraph2024} compresses multiple segment-specific graph indices into a single unified structure, resulting in a more practical overall index size. However, this compactness is achieved by constructing the index according to the sorted order of a chosen relational attribute, inherently limiting support for dynamic vector insertions or attribute updates.
Furthermore,
since SeRF is not inherently designed to support general filtered search,
leveraging it for multi-attribute filtering requires constructing a separate index for each relational attribute.
For example, consider a schema $\mathcal{A}$ with four attributes $a_1$, $a_2$, $a_3$, and $a_4$. If we construct a SeRF for attribute $a_1$, it will only be able to serve queries that impose a predicate $p$ on $a_1$. That is, it cannot serve other predicates such as $a_2 \wedge a_3$.
In order to serve predicates on \emph{any} relational attributes on
$\mathcal{A}$,
we would need to build four SeRF indicies.
This redundancy -- duplicating the vector component once per relational attribute -- leads to prohibitive storage overhead.

DSG~\cite{pengDynamicRangeFilteringApproximate2025} extends SeRF by relaxing the strict ordering constraint, thereby enabling dynamic insertions. Yet, this flexibility comes at the cost of additional space overhead, reintroducing the index size issue that SeRF originally addressed. In our experiments, we observe that the index size of DSG even surpasses that of iRangeGraph.

\subsection{Pre-filtering}
Pre-filtering is a baseline approach for supporting general filtered vector search: it applies all relational predicates to the dataset first and then performs vector search on the resulting filtered subset. While flexible, pre-filtering is only efficient when the combined relational filters yield a extremely selective predicate. The inefficiency stems from the lack of an index over the runtime-generated filtered result, forcing vector search to fall back on a brute-force scan over potentially large intermediate filtered result -- a process that rapidly becomes impractical once the filtered result exceeds a few thousand entries. Consequently, pre-filtering is only effective for predicates with extremely low passrates—typically below 0.1\% for million-scale datasets—where only a handful of vectors remain after filtering.

Moreover, accurately estimating query selectivity with multiple attributes remains a long-standing \emph{cardinality estimation}~\cite{harmouch2017cardinality} challenge, despite advances including recent learning-based approaches~\cite{wang2021are, sun2022learned, kipf2019estimating}. As a result, reliance on pre-filtering introduces high risk of unpredictable latency due to mis-estimation, highlighting the need for a general solution that is less dependent on precise cardinality estimation.

\subsection{Post-filtering}
Post-filtering is another common technique for supporting general filtered vector search. For conjunctive predicates (e.g., $a_1 \land a_2$), a set of $k'$ candidate records is retrieved using vector search as the first step, and then this set is filtered according to the attribute predicate as the second step. However, post-filtering is also fundamentally challenged by the cardinality estimation problem: it is difficult to determine an appropriate initial search size, $k'$, that ensures sufficient candidates will satisfy the later relational filtering. As a result, post-filtering often devolves into multiple search rounds with progressively increasing
$k'$, leading to inefficient and unpredictable performance. This inefficiency is further exacerbated as the predicate’s passrate decreases -- the lower the selectivity, the poorer the performance. This means that more selective predicates can actually increase query latency, contrary to the typical database expectation that query cost should decrease with lower passrates due to less data being accessed.
Nonetheless, post-filtering offers a small advantage over pre-filtering for filtered search: it can leverage any specialized indices built for 1D filtering for vector-search in the first step.

For disjunction predicates (e.g., $a_1 \lor a_2$), the most efficient method is to leverage the pre-built 1D specialized index to locate the eligible records for each queried attribute, union them, and sort the union according to their vector distance to the query. In other words, for this approach, we expect a degradation in QPS when more attributes are involved in a disjunction predicate.

\begin{figure}[t!]
    \centering
    \includegraphics[width=0.8\linewidth]{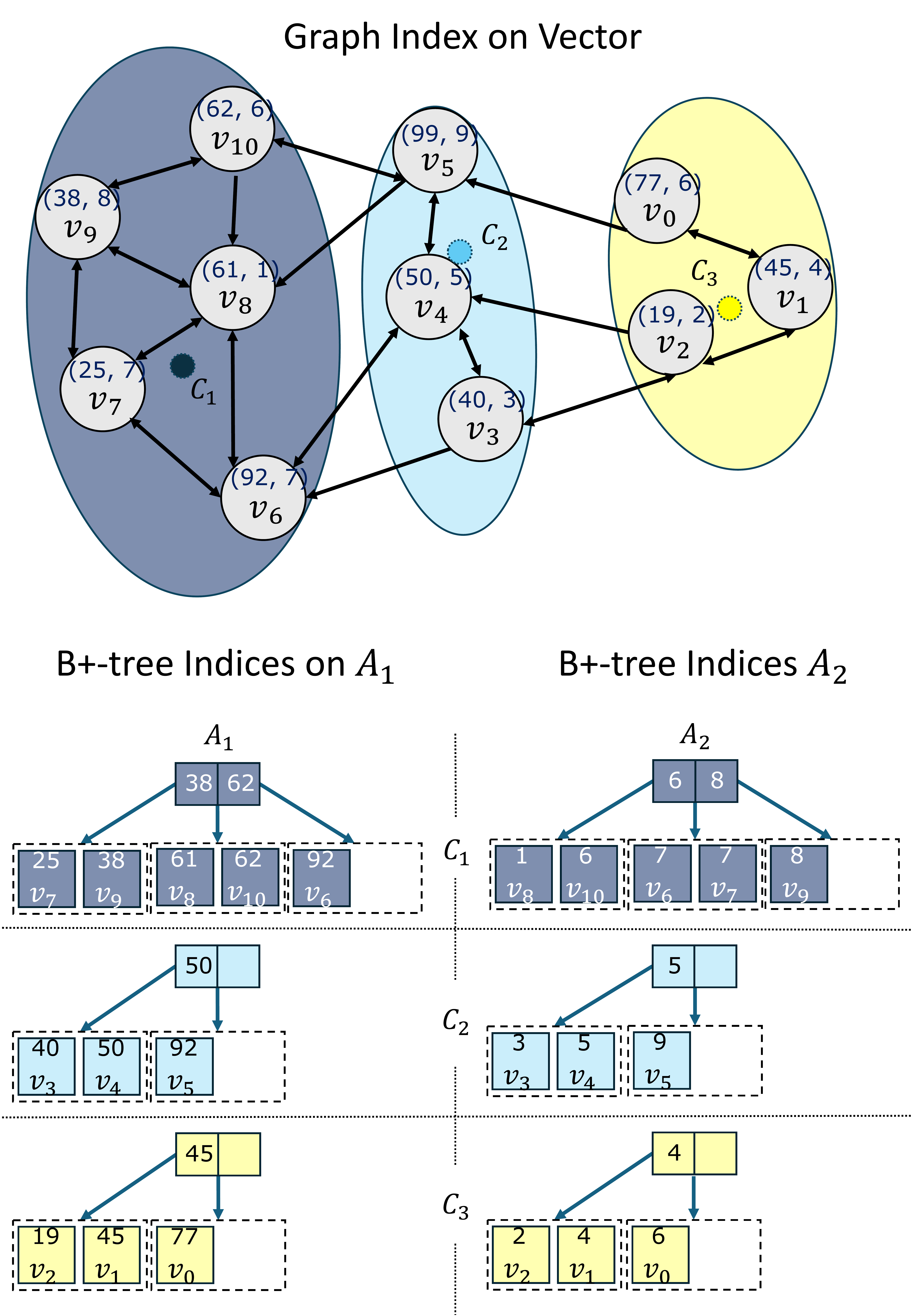}
    \caption{An illustration of \textsc{Compass} index, only the bottom layer graph is shown for HNSW and each color denotes a cluster in the IVF index.}
    \label{fig:compass-overview}
\end{figure}

\subsection{In-filtering and General Filtered Search}
ACORN~\cite{patelACORNPerformantPredicateAgnostic2024} and
NaviX~\cite{sehgalNaviXNativeVector2025}
are among the few general and universal algorithms currently available for supporting filtered vector search without incurring substantial space overhead.
Rather than introducing a novel index, ACORN leverages the widely adopted graph-based index HNSW, applying traversal heuristics and denser vertex neighborhood to restore graph connectivity disrupted by relational filtering.
NaviX improves on ACORN's traversal heuristics without building a denser graph.
For instance, they may explore two-hop neighbors instead of standard one-hop traversal, enabling efficient navigation among eligible records after applying attribute filters. By restricting vector distance computations to records that satisfy the predicate, they limit unnecessary comparisons and prioritize vectors passing the filters.

However, these in-filtering strategies do not always yield high query throughput in practice. The computational savings from reduced distance calculations can be offset by the overhead of locating predicate-passing vectors within the index. As such, while their designs ensure that the number of distance comparisons decreases with lower predicate passrate, this improvement do not necessarily translate into proportional increases in queries-per-second (QPS), due to unavoidable costs in candidate identification and traversal during query processing.

Milvus~\cite{wangMilvusPurposeBuiltVector2021} employs a cost-based model
to pick the best among pre-filtering, in-filtering and post-filtering. Weaviate~\cite{dilockerWeaviate2024} combines only the pre-filtering and the in-filtering.
While functional, their filtered search capabilities are outperformed by specially-optimized designs.
SIEVE~\cite{liSIEVEEffectiveFiltered2025}, as another approach to general filtered search, builds multiple proximity graphs for different filtered subsets within a memory size budget and based on the history query workload.
SIEVE's design is orthogonal to other methods that either build specialized indices~\cite{zuoSeRFSegmentGraph2024, xuIRangeGraphImprovisingRangededicated2024, pengDynamicRangeFilteringApproximate2025, jiangDIGRADynamicGraph2025, zhangEfficientDynamicIndexing2025, liangUNIFYUnifiedIndex2025} or employ special traversal strategies~\cite{patelACORNPerformantPredicateAgnostic2024, sehgalNaviXNativeVector2025}.

\section{The Compass Algorithm}

\subsection{Index Construction}\label{section:construction}

\textsc{Compass} assumes a schema where each record contains a vector and one or more numerical attributes. We build a proximity graph index $\graph$ (HNSW~\cite{malkovEfficientRobustApproximate2020} by default) on the vectors of all records, leveraging the high efficiency of proximity graph for A-$k$NN~\cite{aumullerANNBenchmarksBenchmarkingTool2018, simhadriResultsNeurIPS212022}.
For the numerical attributes,
we first group all records into clusters based on their vectors, using an IVF index like~\cite{jegouProductQuantizationNearest2011}. Then, within each cluster, we build a separate B+-tree for each numerical attribute.
We collectively refer to the combination of IVF and B+-trees as \emph{clustered B+-trees} and denote it as $\clusbtree$.

\begin{example}
{\it
Figure \ref{fig:compass-overview} illustrates an example of the Compass index, where each record consists of one vector and two numerical attributes, $A_1$
  and $A_2$. For instance, vector $v_{10}$ has two attribute values (62, 6) in the figure.
  For the vector component, we construct an HNSW, and only the bottom layer of HNSW is shown.
 For the two relational attributes
 $A_1$ and $A_2$, we first cluster the vectors
 using an IVF and then build indices for
 the relational attributes within each cluster of vectors.
In Figure \ref{fig:compass-overview}, the dataset is partitioned into three clusters,
 $C_1$, $C_2$, and $C_3$ .
 Within each cluster, we build two B+-trees --- one for each attribute.

}
\end{example}

\subsection{Query Processing} \label{sec:search}

\begin{algorithm}[!t]
    \caption{\CompassSearch} \label{algo:compass-search}
    \begin{algorithmic}[1]
        \Require{graph index $\graph$, clustered B+trees $\clusbtree$, query vector $q$, predicate $p$, \#result $k$, expansion factor $\ef$}
        \State{$\sq \leftarrow \text{empty min-heap}$} \label{line:shared-queue-beg}
        \State{$\va \leftarrow \text{empty bitmap}$}
        \State{$\rz \leftarrow \text{empty max-heap}$} \label{line:shared-queue-end}
        \State{$\graph.\GraphOpen(q, p, \sq, \va)$} \label{line:state-init-beg}
        \State{$\clusbtree.\GraphOpen(q, p, \sq, \va)$} \label{line:state-init-end}
        \While{$\text{$\rz$ has not reached size $\ef$}$} \label{line:main-loop-beg}
            \State $records, sel = \graph.\GraphNextFiltered(\sq, \va)$  \label{line:graph-next}
            \For{$record \in records$}
                \State{$\rz.\Push(record) $}
            \EndFor
            \If{$sel < \beta$} \label{line:sel}
                \State{$records = \clusbtree.\ClusBtreeNext(\sq, \va)$} \label{line:clusbtree-next}
                \For{$record \in records$}
                    \State{$\rz.\Push(record) $}
                \EndFor
            \EndIf
        \EndWhile \label{line:main-loop-end}

        \While{$\rz.\Size() > k$} \label{line:pop-rz-beg}
            \State{$\rz.\Pop()$}
        \EndWhile
        \State \Return{$\rz$} \label{line:pop-rz-end}
    \end{algorithmic}
\end{algorithm}

\begin{figure*}[!t]
    \centering
    \includegraphics[width=0.62\linewidth]{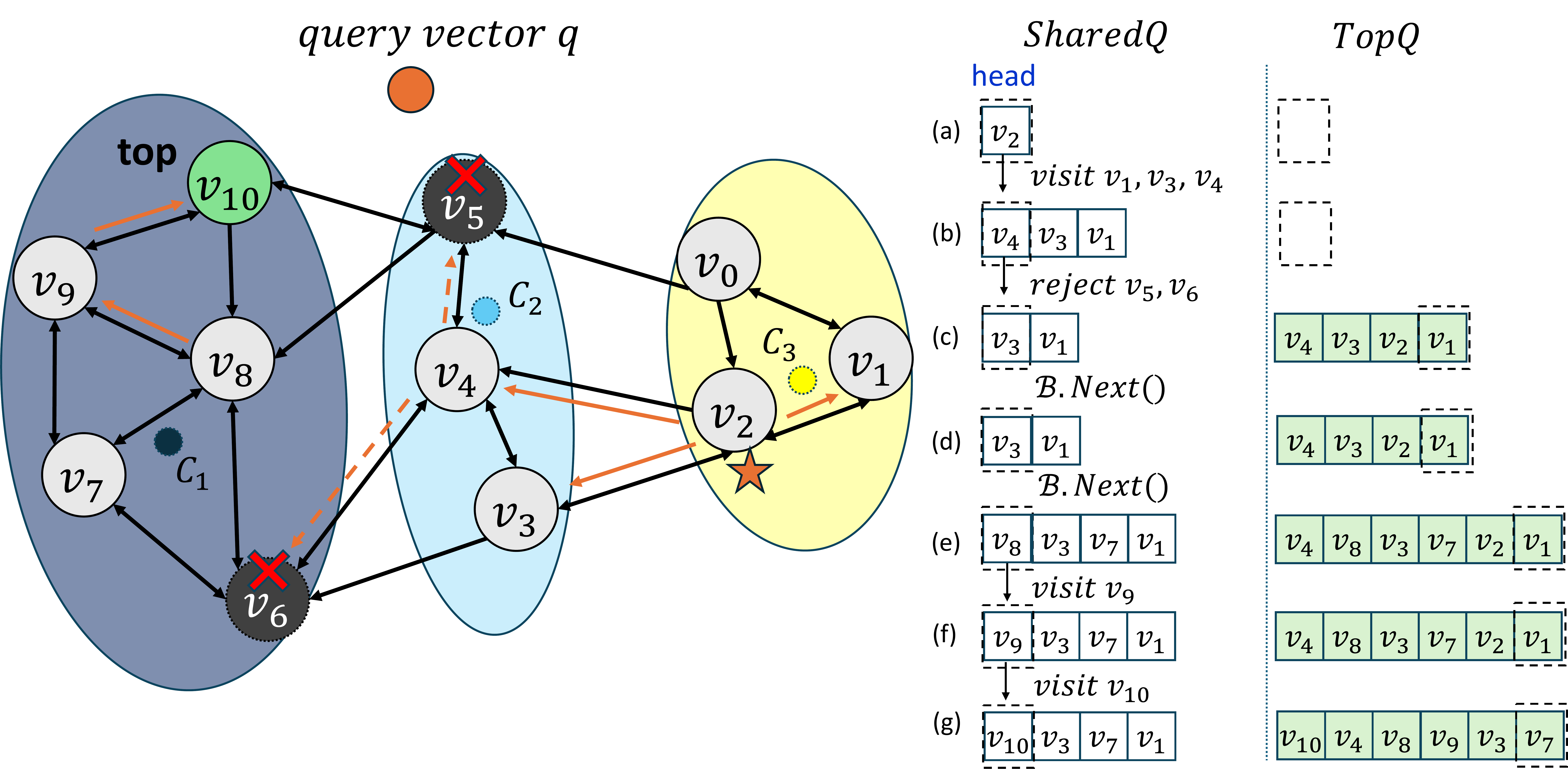}
    \caption{An illustration of \CompassSearch with single attribute $A$. Gray nodes pass the predicate while black ones do not. Green node marks the top-1 result. Orange dot marks the query vector while orange star marks the graph entry point. Cyan, blue and yellow represent different clusters. For conciseness, B+-trees are hidden in the figure.}
    \label{fig:candidate-racing}
\end{figure*}

The key idea of Compass is to jointly leverage the proximity graph $\graph$ (efficient at similarity-based vector search) and the relational indices (efficient at identifying records that satisfy the predicate) for filtered vector search. We employ the proximity graph $\graph$ as the primary driving force given its high efficiency.
However, if only a small number of current candidate's neighbors pass the predicate (i.e., low neighborhood passrate\footnote{We refer to the \textit{neighborhood passrate} of a node in the proximity graph as the portion of its neighbors that satisfy the given predicate.}), the graph traversal can become confined to a component disconnected from other graph regions containing the predicate-satisfying records~\cite{patelACORNPerformantPredicateAgnostic2024}.
To address this, the clustered B+-trees $\clusbtree$ helps the graph traversal escape these isolated components. Specifically, we use $\clusbtree$ to retrieve a batch of predicate-satisfying records from IVF clusters whose centroids are close to the query vector. The proximity graph and clustered B+-trees cooperate via a shared candidate queue that ranks candidates by their vector distance to the query; both indices can contribute to this shared queue.

\Cref{algo:compass-search} details the overall query processing of Compass and formalizes the ideas discussed above. It begins by creating a shared candidate queue, \sq, which maintains the candidate records to visit; a shared visited bitmap, $\va$, which flags the records whose distances have been computed; and the top queue,
\rz, which stores the intermediate query result (Lines~\ref{line:shared-queue-beg} to \ref{line:shared-queue-end}). Then, the query vector, predicate, shared queue and visited bitmap are passed to the proximity graph ($\graph$) and the clustered B+-trees ($\clusbtree$) to initialize their respective search states (Lines~\ref{line:state-init-beg} to~\ref{line:state-init-end}).
Both $\graph$ and $\clusbtree$
follow the pull-based iterator interface~\cite{graefeQueryEvaluationTechniques1993}.
Their \textsc{Open} and \textsc{Next} procedures are detailed in \Cref{algo:graph} and \Cref{algo:clusbtree}, respectively. Currently, we can view the \textsc{Next} interface of $\graph$ as returning a batch of vectors that are encountered during the graph traversal and pass the predicates, and the \textsc{Next} interface of $\clusbtree$ as returning a batch of vectors that pass the predicates and consecutive \textsc{Next} calls return the vectors in their cluster order as discussed earlier.

The main loop (Lines~\ref{line:main-loop-beg} to \ref{line:main-loop-end}) continues until the $\rz$ reaches the preset search size $\ef$ (Line~\ref{line:main-loop-beg}). As such, we can use $\rz$ to control the recall and query processing time.
In the beginning of the loop,
a batch of candidates that pass the predicates are pulled from the proximity graph via $\graph$.\GraphNextFiltered (Line~\ref{line:graph-next}).
Beside returning the candidates,
$\graph$.\GraphNextFiltered
 also returns the
neighborhood passrate $sel$ around the currently visiting candidate (Line~\ref{line:graph-next}).
If $sel$ is lower than a threshold $\beta$ (set to 0.05 by default, \Cref{line:sel}),
the algorithm pulls a batch of candidates that pass the filters
from the clustered B+trees via $\clusbtree$.\ClusBtreeNext (Line~\ref{line:clusbtree-next}).
This injection of candidates mitigates the connectivity issue caused by low passrate, allowing the graph search to continue from these candidates from the clustered B+trees.
This cooperative hand-off is enabled by maintaining the shared candidate queue by both $\graph$ and
$\clusbtree$ during their \GraphNextFiltered operations.
The main loop ends when there are enough candidates in the top queue $\rz$, and by then the top-$k$ result from $\rz$ are returned (Lines~\ref{line:pop-rz-beg} to~\ref{line:pop-rz-end}) as the final search result.

\begin{example}
{\it
\Cref{fig:candidate-racing} illustrates the search process of Compass for an example query on one attribute.
In the figure, nodes colored black represent vectors whose
relational values do not satisfy the predicate, while nodes in gray indicate those that do. The search starts
with using the graph index and selects an entry point $v_2$ and (a) enqueues it into the shared candidate queue \sq.
Since $v_2$'s entire neighborhood passes the predicate, (b) it visits all its neighbors $v_1, v_3, v_4$ and pushes them into \sq.
Next, the search explores $v_4$'s unvisited neighbors $v_5, v_6$ and finds that they all fail the predicate filter. This indicates that the neighborhood selectivity around $v_4$ is poor, indeed 0.
In this case,
the search would (c) consult the clustered B+-trees, which examines
the cluster that is currently closest to the query
($C_2$ in this example),
and use its corresponding B+-tree to retrieve
the predicate-passing records: $v_3, v_4$.

Since $v_3$ and $v_4$
have already been visited,
the clustered B+-trees would examine
the next cluster closest to $q$
($C_1$ in this example) in order to return enough tuples for its \ClusBtreeNext operation.
In the example, it (d) retrieves predicate-passing records from the B+-tree of $C_1$
and adds them into \sq.
If there are too many new candidates,
the clustered B+-trees
would only insert into the \sq and return a sample of them ($v_{7}$ and $v_{8}$). After that, the main loop of \Cref{algo:compass-search}
starts a new iteration and goes back to \Cref{line:graph-next} of \Cref{algo:compass-search}.
Inside $\graph$.\GraphNextFiltered,
it internally (e) first visits $v_8$, then (f) visits $v_9$, and (g) finally reaches the optimal result $v_{10}$.
Since $v_8, v_9$ have already been visited and the only unvisited neighbor $v_5$ of $v_{10}$ fails to pass the predicate,
$\graph$.\GraphNextFiltered in Line~\ref{line:graph-next} of \Cref{algo:compass-search} returns $v_9$ and $v_{10}$ as output.
Finally, since the size of \rz is already large enough in this step,
the search ends by returning the top-$k$ elements in \rz.
}
\end{example}

\subsection{Progressive Search}

\stitle{Motivations} As discussed earlier, we employ the proximity graph as the main driving force in Compass. The search process of proximity graph is controlled by two priority queues, i.e., a min-heap candidate queue that stores the potential records to visit, and a max-heap top queue that maintains the nearest neighbors discovered thus far. The traversal stops when the closest node in the candidate queue is farther from the query than the most distant node in the \efs-sized top queue.
As such, \efs controls the query processing time and result quality.
To utilize proximity graph for filtered search, ACORN~\cite{patelACORNPerformantPredicateAgnostic2024} and NaviX~\cite{sehgalNaviXNativeVector2025} adopt ``in-filtering" by computing distances only for records that pass the predicate.
However, when the predicate's selectivity is low or moderate, the resulting subgraph of predicate-satisfying nodes often becomes disconnected. This leads to a significant performance degradation: the search gets trapped in a local region, wasting computations on nodes that are locally proximate but globally distant from the true, predicate-satisfying nearest neighbors.

\begin{table}[!t]
\centering
\caption{Symbol table of progressive search state}

\resizebox{0.5\textwidth}{!}{
\begin{tabular}{cc}
\hline
Symbol                   & Explanation                                                                            \\ \hline
\cq       & shared min-heap storing next candidate to expand                                       \\
\va       & shared bitmap flagging the visited status of records                                    \\
\rz       & internal max-heap storing visited top records with max size \efs        \\
\recycleq & internal min-heap storing visited records not in internal \rz \\
\result   & internal min-heap storing visited top filtered records from $\graph$                   \\
\efs      & internal expansion factor controlling the search width of $\graph$                                     \\
\deltaefs & step size to increase \efs to enlarge search width                       \\ \hline
\cq       & shared min-heap storing next candidate to expand                                       \\
\va       & shared bitmap flagging the visited status of records                                    \\
\relq     & internal min-heap storing visited top filtered records from $\clusbtree$               \\
\efi      & internal expansion factor of relational indices  \\ \hline             
\end{tabular}
}
\label{table:symbol}
\end{table}

We attribute this ``trapping" problem to using a fixed \efs in existing methods, making traversal carry on without being able to identify the disconnectivity problem. To overcome this, Compass starts with a small initial \efs and progressively enlarges it in discrete steps. In particular, at the end of each step, i.e., when the current \efs limit is reached, the algorithm evaluates its search progress based on the neighborhood passrate. A high neighborhood passrate indicates the graph search is effective and the visited graph region has not become a trap that isolate current candidate from other predicate-satisfying regions; and the algorithm proceeds by enlarging \efs to continue its traversal.
If the neighborhood passrate is low, it signals that the search is likely confined by disconnected subgraph. The algorithm then pivots, querying the clustered B+-trees, which inject new and diverse candidates from which the graph traversal can continue improving, to navigate out of the current local region. After that, \efs can be enlarged accordingly. This technique essentially introduces checkpoints by progressively enlarging \efs, and thus we call it \textit{progressive search}.

To support progressive search, we identify the key variables that describe this process and separately list them for $\graph$ and $\clusbtree$ in~\Cref{table:symbol}. In the following, we describe the $\GraphOpen$ and the $\GraphNext$ interface for both the proximity graph object $\graph$ and clustered B+-trees object $\clusbtree$.

\begin{algorithm}[!t]
    \caption{Proximity Graph's Iteration Interface}
    \label{algo:graph}
    \begin{algorithmic}[1]
        \Function{$\graph$.Open}{$q$, $p$, \sq, \va}
            \State $this.q = q, this.p = p$ \label{line:graph-open-copy}
            \State $this.\cq = \sq$ \label{line:graph-usual-beg}
            \State $this.\va = \va$
            \State $this.\rz \leftarrow \text{empty max-heap}$ \label{line:graph-usual-end}
            \State $this.\result \leftarrow \text{empty min-heap}$ \label{line:graph-extra-beg}
            \State $this.\recycleq \leftarrow \text{empty min-heap}$ \label{line:graph-extra-end}
            \State $\cq.\Push(\SelectEntryPoint(this.q))$ \label{line:graph-usual-entry-point}
        \EndFunction
        \Function{$\graph$.Next}{\sq, \va}
            \State $this.\Call{\ExpandSearch}{\null}$ \label{line:expand-search}
            \While{$\sq$ is not empty}
                \State $\currdist, \currobj = \sq.\Pop()$ \label{line:stop-beg}
                \If{$\currdist > \rz.\Top().dist$} \break \label{line:graph-termination}
                \EndIf \label{line:stop-end}
                \State{\text{$sel =$ the $\currobj$'s neighborhood passrate}} \label{line:stat-sel}
                \If{$sel \ge \alpha$} \label{line:onehop-beg}
                    $\OneHopExpand()$
                \ElsIf{$sel \ge \beta$} \label{line:twohop-beg}
                    $\TwoHopExpand()$
                \Else \label{line:disconnect-beg}
                    $\break$
                \EndIf
            \EndWhile
            \State $records = []$ \label{line:graph-next-res-beg}
            \While {\text{$\result.\NotEmpty()$ and $\cnt\text{++} < k$}}
                \State{$records.\Push(\result.\Pop())$}
            \EndWhile
            \State \Return $records, sel$ \label{line:graph-next-res-end}
        \EndFunction

        \Function{$\graph$.\ExpandSearch}{\null}
            \State \text{$this.\efs$ += $this.\deltaefs$} \label{line:graph-next-expand}
            \While{\text{$\recycleq.\NotEmpty()$ and $\rz.\Size() < this.\efs$}} \label{line:graph-next-bk-beg}
                \State $top = \recycleq.\Pop()$
                \State $\rz.\Push(top)$
                \If{\text{$top$ never added to $\sq$}}
                    \State $\sq.\Push(top)$
                    \If{\text{$p(top)$ is true}}
                        \State $\result.\Push(top)$
                    \EndIf
                \EndIf
            \EndWhile \label{line:graph-next-bk-end}
        \EndFunction
    \end{algorithmic}
\end{algorithm}

\stitle{Operations on the proximity graph}
\Cref{algo:graph} lists the $\GraphOpen$ and $\GraphNext$ procedures for the proximity graph object $\graph$. In particular, $\graph.\GraphOpen$ begins by referencing the query vector and predicate. (Line~\ref{line:graph-open-copy}). It then references the shared queue $\sq$ as well as the shared visited bitmap $\va$, and initializes the internal top queue $this.\rz$ (marked with \textit{this} to differentiate with the global top queue in the main loop) like in standard proximity graph search, with the key exception that its candidate queue and visited bitmap are shared with the clustered B+-trees (Line~\ref{line:graph-usual-beg} to~\ref{line:graph-usual-end}).
The graph index also internally maintains its own result queue $\result$ to store the filtered results and recycle queue $\recycleq$ to support the \GraphNext interface (Line~\ref{line:graph-extra-beg} to~\ref{line:graph-extra-end}). Finally, it finds the entry point
and pushes it to the shared candidate queue like in standard proximity graph search (Line~\ref{line:graph-usual-entry-point}).

The graph search begins by enlarging its search size \efs by \deltaefs, enabling it to continue from where it left off in previous step (Line~\ref{line:expand-search}). It then pops the best candidate from the shared queue and checks the stop condition like standard proximity graph search (Line~\ref{line:stop-beg} to~\ref{line:stop-end}).

For the popped candidate, the search employs an adaptive expansion strategy based on its neighborhood's predicate passrate (Line~\ref{line:stat-sel}). If the passrate is moderately large ($\ge \alpha$, with $\alpha$ set to 0.3 by default), it opts for a one-hop expansion (Line~\ref{line:onehop-beg}), i.e.,  visiting all the unvisited one-hop neighbors of the current candidate. If the passrate is moderately low ($\ge \beta$ but $< \alpha$, with $\beta$ set to 0.05 by default), it employs a limited two-hop expansion (Line~\ref{line:twohop-beg}), i.e., visiting the unvisited predicate-passing one-hop neighbors as well as a subset of unvisited predicate-passing two-hop neighbors of the current candidate. The rationale is that two-hop neighbor expansion leads to predicate-passing records outside the neighborhood. We visit only a subset of predicate-passing two-hop neighbors to avoid excessive attribute filtering.

If the passrate is extremely low (i.e., $<\beta$), the proximity graph determines it is disconnected from other predicate-passing regions
and prepares to consult the clustered B+-trees for connectivity enhancement (Line~\ref{line:disconnect-beg}). Finally, the close, predicate-passing records found in this round are returned (Line~\ref{line:graph-next-res-beg} to~\ref{line:graph-next-res-end}).
The graph search's termination is dynamically determined (Line~\ref{line:graph-termination}). Furthermore, because the graph index is predicate-agnostic, the number of predicate-passing records found can vary between rounds. Therefore, our $\GraphNext$ function returns a batch of results, rather than a single-item iterator (cf.~\cite{graefeQueryEvaluationTechniques1993}).

To detail the \ExpandSearch mechanism, note that the parameter $\efs$ sets the graph search width. At the beginning of $\graph$.\ExpandSearch, this parameter is incremented, which semantically enlarges the search width (Line~\ref{line:graph-next-expand}). To materially execute this expansion, a ``recycle queue" is employed. This queue maintains the intermediate visited records, which are used to set the candidate queue and top queue to the precise state they would have been in if the search width had been this large from the start (Line~\ref{line:graph-next-bk-beg} to~\ref{line:graph-next-bk-end}).

\begin{algorithm}[!t]
    \caption{Clustered B+-trees' Iteration Interface} \label{algo:clusbtree}
    \begin{algorithmic}[1]
        \Function{$\clusbtree$.Open}{$q$, $p$, \sq, \va}
            \State $this.q = q, this.p = p$ \label{line:clusbtree-open-init-beg}
            \State $this.\bg = nil, this.\ed = nil$ \label{line:clusbtree-open-init-end}
            \State $this.\cq = \sq$ \label{line:clusbtree-q-beg}
            \State $this.\va = \va$
            \State $this.\relq \leftarrow \text{empty min-heap}$ \label{line:clusbtree-q-end}
            \State $this.\cg.\GraphOpen(q, \textbf{TRUE})$ \label{line:cluster-graph-open}
        \EndFunction
        \Function{$\clusbtree$.Next}{\sq, \va}
            \State $\cnt = 0$ \label{line:clusbtree-propose-beg}
            \While{$\cnt < \efi$} \label{line:clusbtree-expansion}
                \If{\text{$\bg$ reached $\ed$}} \label{line:btree-query-beg}
                    \State{$clusidx = \cg.\GraphNext()$} \label{line:cluster-graph-next}
                    \State{$\bg, \ed = \indices[clusidx].\Search(p)$}
                \EndIf \label{line:btree-query-end}
                \If{$!\va[*beg]$}
                    \State $\relq.\Push(\{\DistFunc(q, *beg),*beg\})$
                    \State $\va[*beg] = true$
                    \State $\cnt\text{++}$
                \EndIf
                \State $\bg$++
            \EndWhile \label{line:clusbtree-propose-end}
            \State{$\cnt = 0, records = []$} \label{line:clusbtree-cleanup-beg}
            \While{\text{$\relq.\NotEmpty()$ and $\cnt\text{++} < k/2$}} \label{line:clusbtree-batchsz}
                \State{$\sq.\Push(\relq.\Top())$}
                \State{$records.\Push(\relq.\Pop())$}
            \EndWhile
            \State \Return{$records$} \label{line:clusbtree-cleanup-end}
        \EndFunction
    \end{algorithmic}
\end{algorithm}

\stitle{Operations on clustered B+-trees} When the graph traversal become trapped at local region due to low passrate and poor connectivity, Compass pivots to the clustered B+-trees to inject new candidates to navigate the graph traversal out of the local region. The central challenge then becomes efficiently selecting the closest clusters to probe for predicate-passing records while keeping the selection overhead low.

A straightforward solution is to utilize a linear scan over all cluster centroids like a standard IVF. The computation cost is high as there are usually many centroids. Moreover, we seldom need to query the predicate-passing records from all the clusters, and thus the solution wastes computation.
To reduce computation, an alternative is to probe a pre-determined number of closest clusters,  \textit{nprobe}, via a separate approximate similarity search on the centroids (e.g., with a proximity graph on the centroids).
This solution requires difficult parameter tuning: a conservative \textit{nprobe} cannot inject a sufficient number of new candidates to navigate out of the local region, while an aggressive \textit{nprobe} incurs superfluous computational overhead on clusters that do not contribute to the final search results.

To resolve the problem, we propose a more dynamic, ``on-demand" cluster ranking strategy. In particular, we build a proximity graph on the cluster centroids, named as cluster graph $\cg$, and reuse the previous progressive search method to fetch close clusters.
Each time $\clusbtree$.\ClusBtreeNext is called, the \textit{efs'} for searching the cluster graph is similarly incremented {by \deltaefs'} to obtain {\deltaefs'} more clusters. Such a design avoids both the exhaustive computation of a full ranking and the ad-hoc nature of a fixed \textit{nprobe} heuristic, while intrinsically balancing computational efficiency with the required candidate sufficiency.

\Cref{algo:clusbtree} lists the $\ClusBtreeOpen$ and $\ClusBtreeNext$ procedures for the clustered B+-trees object $\clusbtree$. At the beginning of $\clusbtree$.\ClusBtreeOpen, the query vector and predicate are referenced, and the relational iterators are initialized (Line~\ref{line:clusbtree-open-init-beg} to~\ref{line:clusbtree-open-init-end}). It then references the shared candidate queue $\sq$, shared visited bitmap $\va$ and initializes its own internal ``relational queue" for storing close, predicate-passing candidates (Line~\ref{line:clusbtree-q-beg} to~\ref{line:clusbtree-q-end}).
Clustered B+trees maintains the small cluster graph ($\cg$) built on the cluster centroids to progressively retrieve close clusters. Since this cluster graph's purpose is to find centroids by vector proximity, it is passed with an ``always-true" predicate, causing it to degenerate into a pure progressive similarity search.
$\clusbtree.\GraphOpen$ concludes by initializing the search state for the cluster graph without sharing candidate queue or visited bitmap (Line~\ref{line:cluster-graph-open}).

When the clustered B+-trees are invoked to propose candidates via $\clusbtree$.\ClusBtreeNext, it fetches a fixed number, $\efi$, of predicate-passing records from the close clusters by querying relational indices inside each cluster (Line~\ref{line:clusbtree-propose-beg} to~\ref{line:clusbtree-propose-end}).
The expansion factor $\efi$ is analogous to the proximity graph search's expansion factor $\efs$ (Line~\ref{line:clusbtree-expansion}). Both parameters ensure that each component performs more work (e.g., distance computations) than the number of results returned in a single batch to return quality close records.
If current cluster does not contain sufficient number of records, a new cluster is pulled from the cluster graph $\cg$ to continue the relational candidate proposal (Line~\ref{line:btree-query-beg} to~\ref{line:btree-query-end}).
We note that cluster graph does not share candidate queue or visited bitmap by explicitly omitting them in the function arguments (Line~\ref{line:cluster-graph-next}).
Finally, top filtered records are pushed to the shared queue for potential neighbor expansion, and are returned as result (Line~\ref{line:clusbtree-cleanup-beg} to~\ref{line:clusbtree-cleanup-end}). The $k/2$ batch size (Line~\ref{line:clusbtree-batchsz}) is chosen to accommodate the potentially varied number of records (from $0$ to $k$) returned by the proximity graph search $\graph.\GraphNext$.

\stitle{Details} There are several details omitted from the discussion of the algorithms.
First, the proximity graph $\graph$ and the clustered B+-trees $\clusbtree$ share a common bitmap to track the visited status of all records, ensuring that the vector distance for any given record is computed only once.
Second, during graph traversal, the ``visit" to a record, as detailed in \Cref{algo:visit}, entails computing the record's vector distance and updating the corresponding queues, serving the purpose of supporting progressive search. Particularly, other than flagging visited, computing distance and maintaining the $\sq$ and $\rz$ like in a standard HNSW (Line~\ref{line:visit-standard-beg} to~\ref{line:visit-standard-end}), the record is further pushed into result queue to be returned as filtered close record if it passes the predicate (Line~\ref{line:to-result-queue-beg} to~\ref{line:to-result-queue-end}). If top queue is full and the record is not close enough to the query vector, it is pushed into the recycle queue to be popped out potentially in future step (Line~\ref{line:to-recycle-beg} to~\ref{line:to-recycle-end}).

\begin{algorithm}[!t]
    \caption{\MaintainQueues}
    \label{algo:visit}
    \begin{algorithmic}[1]
        \Require{unvisited $\currobj$, passing predicate or not $\passed$}
        \State $\va[record] = true$ \label{line:visit-standard-beg}
        \State $\currdist = \DistFunc(q, \currobj)$
        \If{$\rz.\Size() < \efs$ or $\currdist < \rz.\Top().dist$}
            \State $\sq.\Push(\{\currdist, \currobj\})$
            \State $\rz.\Push(\{\currdist, \currobj\})$
            \If{$\rz.\Size() \ge \efs$}
                \State $\recycleq.\Push(\rz.\Pop())$
            \EndIf \label{line:visit-standard-end}
            \If{$\passed$} \label{line:to-result-queue-beg}
                \State $\result.\Push(\{\currdist, \currobj\})$
            \EndIf \label{line:to-result-queue-end}
        \Else \label{line:to-recycle-beg}
            \State $\recycleq.\Push(\{\currdist, \currobj\}$
        \EndIf \label{line:to-recycle-end}
    \end{algorithmic}
\end{algorithm}

\subsection{Discussions}\label{section:discussions}

By combining a proximity graph with IVF-enhanced relational indices, \textsc{Compass} benefits from the following advantages.

\stitle{Generality} First and most importantly, Compass generalizes across different numbers and types of attribute filters, tackling the general filtered vector search problem.
This is because Compass' proximity graph refers to attribute information on demand only during the index search, instead of being influenced by the attribute information during the index construction.

This is a stark difference from existing specialized indices, e.g., SeRF~\cite{zuoSeRFSegmentGraph2024} and iRangeGraph~\cite{xuIRangeGraphImprovisingRangededicated2024}, that modify the underlying proximity graph structure to support a limited number of attribute filter (indeed 1 numerical attribute filter).
When there is update on the attribute value, these methods need to completely rebuild the index from scratch. While in our case, only the B+-trees need to be updated with a small overhead.
We note that supporting general filter is important because our industry collaborator handles tens of attributes and arbitrary conjunctions and disjunctions over the filters on individual attributes, forming extremely complex predicates.

\stitle{Efficiency} As will be shown in our experiments, Compass maintains a reasonably high query processing efficiency across predicate patterns (e.g. single- and multi-attributes, varying passrates, conjunction, disjunction).
In particular, when the passrate is high or moderate, ``in-filtering'' traversal is efficient by leveraging the graph connectivity.
In this case, Compass will seldom engage the clustered B+-trees and mainly employ the proximity graph.
When the passrate is low, pulling from the clustered B+-trees can supply quality candidates.
In this case, Compass will mainly rely on the clustered B+-trees to identify the predicate-passing records in the order of their cluster centroid distance to the query.
As a result,
we expect the QPS of Compass increases with more selective queries
because the clustered B+-trees
are (1) unaffected by the graph disconnections (as they are not graph-based);
and (2)  benefit from having fewer candidates, resulting in less work.
Compass smoothens the
transition between the two cases by using a shared candidate queue between the proximity graph and clustered B+-trees with the neighborhood passrate as the signal to pivot in between.

Compass is also efficient in the index construction and storage. The IVF index and the relational indices can be built quickly, 
in comparison to specialized indices like SeRF~\cite{zuoSeRFSegmentGraph2024}, ACORN~\cite{patelACORNPerformantPredicateAgnostic2024}, iRangeGraph~\cite{xuIRangeGraphImprovisingRangededicated2024} and DSG~\cite{pengDynamicRangeFilteringApproximate2025} that incur significantly longer time for index construction.

In terms of storage, the clustered B+-trees store the cluster centroids, edges of a small cluster graph and relational indices. These overhead are small compared to the storage required for the proximity graph's edge information. Overall, Compass introduces only a minor storage overhead on top of the base proximity graph index, in comparison to specialized indices that would require one index per attribute in multi-attribute setting.

\stitle{Flexibility} By separating the indices for vector similarity and attribute filtering, Compass benefits from the flexibility in index choice.
For instance, the HNSW index can be replaced with a different proximity graph algorithm like NSG~\cite{fuFastApproximateNearest2019}, or the per-attribute B+-trees could be replaced with a single
multi-dimensional tree like
R-tree~\cite{guttmanRtreesDynamicIndex1984}.
This allows Compass to seamlessly integrate with the latest development in indexing technique.
For example, the update on vectors (e.g., insertion or deletion) can be easily supported by recently-developed algorithms to update the proximity graph index~\cite{liuWolverineHighlyEfficient2025} and the IVF index~\cite{xuSPFreshIncrementalInPlace2023} respectively.
Besides, update to the relational attributes only requires update to relational indices, leaving the proximity graph and IVF clusters intact.

\stitle{Limitations} The highly-modular and general-filter design of \textsc{Compass} inevitably influences its search performance when compared to highly-specialized indices in their optimal settings (specifically in single-attribute case).
However, as demonstrated in our experiments, the performance gap between Compass and these specialized indices in their preferred settings is often minimal, and in some cases, Compass even outperforms them.
Additionally, it's important to highlight that these specialized indices, when applied to a relational schema with $n$ attributes, require $n$ times the storage redundancy for the vector component.

Currently, when querying the predicate in the clustered B+-trees, \textsc{Compass} selects the B+-tree on one attribute as the access path and conducts a linear scan to filter on the remaining attributes over the returned records on that selected attribute.
As an engineering improvement,
we can treat \emph{records in each cluster $C_i$ as an independent table $T_{C_i}$} in an RDBMS. When Compass needs candidate records from $C_i$, it can delegate the relational predicate $p$ to the underlying RDBMS as a standard query $Q_p$ on $T_{C_i}$, allowing any mature query optimizer to automatically select the best access plan, further reducing the overall query latency. 
Industrial systems such as MySQL/InnoDB places no practical limit on the number of tables and allows a maximum of 64 indices per table. Excessive overhead about too many tables/indexes has rarely been reported except at extreme scales (e.g. 100,000 tables). \Cref{section:evaluated-methods-and-index-config} will show that Compass uses a reasonable number of clusters, so that it can be conveniently integrated into databases in future.

\section{Evaluation}
In this section, we evaluate the performance of Compass against a range of existing methods on various datasets.

\subsection{Datasets, Workloads, and Metrics}

\begin{table}[!t]
\centering
\caption{Dataset specification.}

\begin{tabular}{@{}cccc@{}}
\toprule
Dataset  & \#Vectors & \#Dimensions & Type            \\ \midrule
CRAWL    & 1,989,995 & 300         & text embedding    \\
GIST     & 982,694    & 960         & image descriptor \\
VIDEO    & 1,000,000 & 1024        & video embedding  \\ 
GLOVE100 & 1,183,514 & 100         & word embedding   \\
\bottomrule
\end{tabular}
\label{table:datasets}
\end{table}

\begin{table*}[t!]
\centering
\caption{Comparing the index sizes of \textbf{Compass} with baselines.}

\resizebox{\textwidth}{!}{
\begin{tabular}{@{}ccccccccc@{}}
\toprule
\textbf{Dataset}     & \textbf{Compass}                                      & \textbf{SeRF}        & \textbf{NaviX}       & \textbf{{ACORN}} & \textbf{{Milvus$^*$}}      & \textbf{{Weaviate$^\dagger$}}    & \textbf{iRangeGraph} & \textbf{DSG}         \\
\multicolumn{1}{l}{} & \multicolumn{1}{l}{(Graph + IVF + B+-trees)} & \multicolumn{1}{l}{} & \multicolumn{1}{l}{} & \multicolumn{1}{l}{}                   & \multicolumn{1}{l}{} & \multicolumn{1}{l}{} & \multicolumn{1}{l}{} & \multicolumn{1}{l}{} \\ \midrule
CRAWL                & 275+13+48*4=480MiB                                    & 269*4=1076MiB        & 592MiB               & 1.46GiB        & 500MiB     & N/A                  & 1.8*4=7.2GiB         & 3.7*4=14.8GiB        \\
GIST                 & 138+38+24*4=272MiB                                    & 150*4=600MiB         & 915MiB               & 736MiB         & 100MiB    & N/A                  & 1.1*4=4.4GiB         & 2.4*4=9.6GiB         \\
VIDEO                & 137+80+24*4=313MiB                                    & 129*4=516MiB         & 999MiB               & 828MiB         & 300MiB     & N/A                  & 0.7*4=2.8GiB         & 0.9*4=3.6GiB         \\
GLOVE100             & 164+9.3+24*4=269.3MiB                                 & 139*4=556MiB         & 135MiB               & 983MiB         & 300MiB    & N/A                  & 0.9*4=3.6GiB         & 1.6*4=6.4GiB         \\ \bottomrule
\end{tabular}
}
${}^*$ No function to provide index sizes.  Estimated according to https://milvus.io/tools/sizing. 
\\${}^\dagger$ No function to provide index sizes. Not feasible to infer the index sizes because of its compact file storage.

\label{table:size}
\end{table*}

We evaluate on four vector datasets: CRAWL, GIST, VIDEO, and GLOVE100.
CRAWL\footnote{https://commoncrawl.org} consists of 300-dimensional text embeddings~\cite{fuHighDimensionalSimilarity2022} derived from crawled web content.
GIST\footnote{http://corpus-texmex.irisa.fr} comprises 960-dimensional floating-point image feature descriptors.
VIDEO\footnote{https://research.google.com/youtube8m} contains 1,024-dimensional video feature vectors subsampled from the YouTube-8M dataset.
GLOVE100\footnote{https://nlp.stanford.edu/projects/glove} contains 100D word embeddings obtained from GLoVe algorithm~\cite{penningtonGloVeGlobalVectors2014}.
They span a variety of source data modalities and numbers of dimensions.
These datasets have their query vectors given and separated from the base vectors.
The number of base vectors and the dimension of the base vector for every dataset is detailed in \Cref{table:datasets}. We note that GIST and VIDEO contain duplicate vectors and we have deduplicated them before all evaluations.
For each vector, we augment it with four uniformly generated relational attributes.

By default, each query is a general range-filtered query with \textit{k}=10 and a selectivity (passrate) of 30\% for each relational attribute, achieved by appropriately adjusting the query range. Each experiment runs a workload of 200 queries, focusing solely on search operations, with no insertions or deletions.

Following existing works~\cite{patelACORNPerformantPredicateAgnostic2024, engelsApproximateNearestNeighbor2024, xuIRangeGraphImprovisingRangededicated2024, zuoSeRFSegmentGraph2024}, we measure the average throughput in the unit of queries per second (QPS) and measure the average accuracy using recall defined as $|\mathcal{R} \cap \mathcal{R}'| / k$, where $\mathcal{R}$ is the result set and $\mathcal{R}'$ is the groundtruth set, supposing all the queries can return up to \textit{k}=10 nearest vectors.
Additionally, we track the number of vector distance computations (\#Comp). For Compass, \#Comp includes the distance computations incurred on clustered B+-trees as well as those incurred on proximity graph.

\noindent \textbf{Platform and Configuration}. All our code is implemented with \text{C++}. All the methods are compiled with GCC version 10.2.1 and compilation option \texttt{-O3 -march=native}. SIMD instructions have been enabled for all compared methods. All the experiments are conducted on Debian 11 with Intel(R) Xeon(R) CPU E5-2620 v4 @ 2.10GHz and 256GB of RAM. Index search performance is evaluated with single thread.

\subsection{Evaluated Methods and Index Size}\label{section:evaluated-methods-and-index-config}
We mainly compare with the following existing works.

\squishlist
    \item \uline{SeRF}~\cite{zuoSeRFSegmentGraph2024}.
    SeRF represents the state-of-the-art in specialized indexing for 1D attribute filtering. When evaluating with multiple attributes,
    we build a 1D-specialized index for each relational attribute and use post-filtering for conjunction.
    Following \cite{zuoSeRFSegmentGraph2024} and based on grid search, we set the construction expansion factor {\efc}=200 across all datasets, set maximum out degree \textit{M}=32 on CRAWL and GIST and \textit{M}=64 on VIDEO and GLOVE100 for SeRF. For its specific index construction parameter, i.e. \textit{efmax}, we use its default value 500.

    Table \ref{table:size} reports the size of SeRF under this configuration alongside other methods.
    For all 1D-specialized indices (SeRF, iRangeGraph \cite{xuIRangeGraphImprovisingRangededicated2024}, DSG \cite{pengDynamicRangeFilteringApproximate2025}),
    we have to build four of them, one per relational attribute.

    Based on their sizes,
    we exclude SeRF’s successor, DSG~\cite{pengDynamicRangeFilteringApproximate2025}, from our performance study: under our experimental setting, which does not involve vector insertions, DSG would offer no performance benefit over SeRF but incurs substantially greater space overhead.
    For the same reason, we exclude iRangeGraph~\cite{xuIRangeGraphImprovisingRangededicated2024} from our performance study as its index size is almost an order larger than SeRF and Compass.
Any observed performance advantage of iRangeGraph would therefore reflect a trade-off between speed gains and the unreasonable expense of memory bloat.

\item \uline{NaviX}~\cite{sehgalNaviXNativeVector2025} and \uline{ACORN}~\cite{patelACORNPerformantPredicateAgnostic2024} are, to date, the only two solutions specifically designed for general filtered vector search.
\uline{Milvus}~\cite{wangMilvusPurposeBuiltVector2021} and \uline{Weaviate}~\cite{dilockerWeaviate2024} integrate basic pre-filtering, in-filtering and post-filtering mechanisms within their engines.

Milvus and Weaviate employ plain HNSW as its index.
In consistence with SeRF, we set the construction expansion factor \textit{K}=200, set maximum out degree \textit{M}=16 on CRAWL and GIST, \textit{M}=32 on VIDEO and GLOVE100, since the out degree of bottom-level HNSW graph doubles that amount. We set the same parameters \textit{M} and \textit{K} for ACORN. For ACORN-specific parameters, following the recommendation in~\cite{patelACORNPerformantPredicateAgnostic2024}, we set $\beta=64$ and set $\gamma = 100$ because the smallest passrate will be $\approx 1\%$.

\item \uline{Prefiltering} is a basic strategy for exact general filtered search. It filters the records with the relational index and then 
compares each filtered record with the query vector.

\squishend

For Compass, its construction involves HNSW building, K-mean clustering and B+tree construction.
We set the expansion factor during construction \textit{K}=200; set maximum out degree \textit{M}=16, number of clusters \textit{nlist}=10000, cluster graph maximum out degree \textit{M'}=4 on CRAWL and GIST; \textit{M}=32, \textit{nlist}=20000, \textit{M'}=8 on VIDEO and GLOVE100.
\textit{K, M, M'} follow aforementioned settings or existing practices. \textit{nlist} is determined using Elbow method. For B+-trees, we use the default fanout factor 64 from the employed library.
As all methods' query execution is controlled by the expansion factor $\ef$, we vary \ef from 10 to 1000, incremented by 5 before 100, by 10 before 200, by 50 before 500, by 100 before 1000.
For Compass' specific search parameters, we fix $\deltaefs$=$k$ for the proximity graph $\graph$; we set $\deltaefs'$=20 for the cluster graph $\cg$ on all datasets; we initialize the internal $\efs = \deltaefs, \efs'=\deltaefs'$ for convenience.
We set $\efi$=50 for CRAWL, GIST, and VIDEO, and $\efi$=100 for GLOVE100.

Now, we take a closer look at the index sizes of Compass, SeRF, NaviX and ACORN
in Table~\ref{table:size}.

Compass, as a general-purpose solution, maintains three complementary structures:
(1) a vanilla HNSW graph index to store the neighbor IDs of base vectors;
(2) IVF centroids together with a small cluster graph ; and
(3) a B+-tree for each relational attribute within each cluster.
Unlike specialized 1D indexing with post-filtering,
Compass requires no vector index duplications across
relational attributes,
its index size is about 50\% of SeRF, 30\% of ACORN, 5\% of iRangeGraph,
and 2.5\% of DSG.
Notably, Compass’s index size could be further reduced by replacing the B+-trees with learned indexes  \cite{gre}, and even more so by leveraging the static nature of the dataset—since no vector insertions occur—making it possible to employ static learned indices like
PGM \cite{pgm}, which are even more compact.
NaviX, as a general-purpose solution like Compass, exhibits index sizes that are comparable to those of Compass. ACORN may incur larger index storage because it builds a denser graph than Compass and NaviX.

\subsection{Conjunctions}\label{section:conjunctions}

\begin{figure}[!t]
    \centering
    \includegraphics[width=\linewidth]{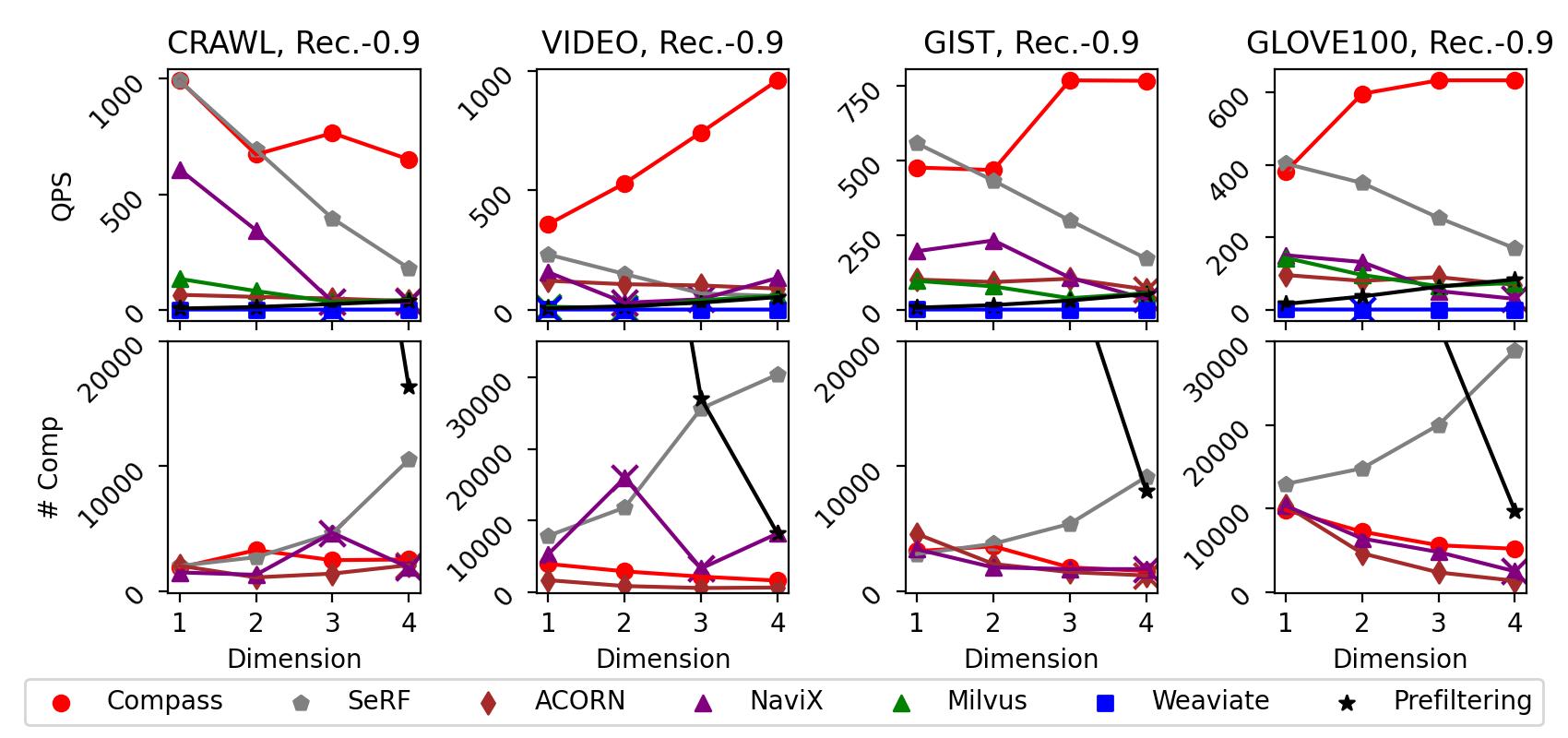}
    \caption{Conjunction Range Filtering. 0.9 Recall. }
    \label{fig:2d-recall-0.9}
\end{figure}

\begin{figure}[!t]
    \centering
    \includegraphics[width=\linewidth]{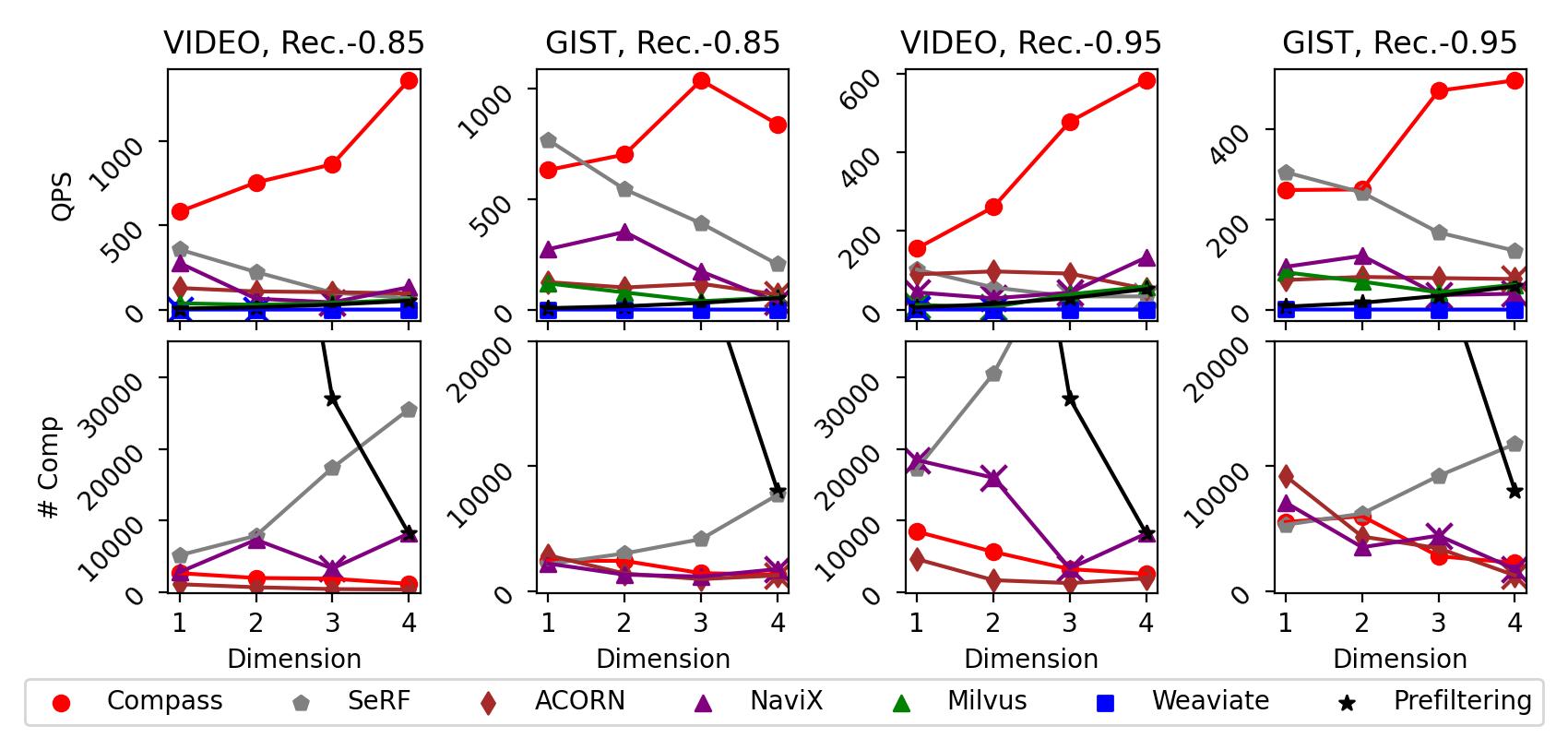}
    \caption{Conjunction Range Filtering. 0.85/0.95 Recall.}
    \label{fig:2d-recall-0.85-0.95}
\end{figure}

\Cref{fig:2d-recall-0.9} and \Cref{fig:2d-recall-0.85-0.95}
present the query throughput (queries per second, QPS) and the number of vector distance computations for Compass, SeRF (with post-filtering), NaviX, ACORN, Milvus, Weaviate, and Prefiltering as the number of conjunctive relational predicates varies from one to four.
Each attribute forms a part of a conjunctive predicate, with experiments conducted under three recall thresholds: 0.85, 0.9 and 0.95.
\Cref{fig:2d-recall-0.9}
presents the full results of recall 0.9 on all four datasets.
\Cref{fig:2d-recall-0.85-0.95} presents the results of recall 0.85 and 0.95 on VIDEO and GIST only
due to space reasons (results on the other two datasets are similar).

Since we set the selectivity (passrate) of each attribute to 30\%, the overall passrate for the conjunctive predicate decreases multiplicatively --
from 30\% with one attribute, to $0.3^2 \approx 10$\% for two attributes,
$0.3^3 \approx 3$\% for three attributes,
and $0.3^4 \approx 1$\%
 for four attributes. These scenarios reflect practical settings, where range predicates in traditional databases typically span moderately selective (30\% passrate) to highly selective (1\% passrate) queries~\cite{selective1, selective2}.

The results are consistent across all recall thresholds: Compass achieves QPS comparable to SeRF (with post-filtering) in low-dimensional scenarios (1D and 2D) and outperforms SeRF in higher dimensions (3D and 4D). In all cases, Compass consistently surpasses NaviX, ACORN and remaining baselines in QPS performance.

We observe that NaviX and ACORN generally incurs fewer distance computations than Compass but ultimately achieves a lower QPS. This is because they only calculate vector distances for predicate-passing records. However, to maintain navigation through the graph despite potential disruptions caused by attribute filtering, they need to frequently visit two-hop neighbors. As a result, substantial time is spent on predicate evaluation for a quadratic number of neighbors, which severely hurts their overall QPS. ACORN maintains an even larger size of neighborhood for each record, resulting in lower QPS than NaviX in many cases.

We also note that NaviX often fails to reach the target recall, even when the number of computations is small. These cases are marked with ``$\times$'' labels in the figures. In these figures, the QPS and the number of computations for NaviX marked with $\times$ represent the results after it has exhausted the largest search size (\textit{ef}=1000). The reason NaviX frequently fails to achieve the target recall is that exploring two-hop neighbors is ineffective when the graph is disconnected into disjoint components due to attribute filtering. In such cases,
any graph traversal will remain trapped within a component indefinitely.

For Milvus and Weaviate, as they have no interface to expose their number of distance computations, we only report their QPS in all the evaluations.
Milvus adopts a cost-based method on the basic pre-filtering, in-filtering and post-filtering strategy; while Weaviate combines the pre-filtering and in-filtering strategy.
The lack of filtered-search optimizations makes them generally under-perform to other methods.

For Prefiltering, it is not competitive unless the number of conjunctions increases to four, resulting in an overall passrate decreasing to approximately 1\%. 
This is because, when dealing with large datasets, a selectivity as stringent as 1\% still generates too many intermediate results after filtering. As an exact method, each of these results requires a distance comparison with the query vector, which places it at the lower end of the QPS spectrum.

Another observation is that Compass exhibits a desirable property similar to that of classical relational database systems, where QPS generally increases as the number of relational filters grows. 
In contrast, the other methods
perceive additional relational filters as increasingly challenging due to the increased graph disconnectivity. They experience the opposite trend -- QPS degrades as the number of relational attributes increases.

\begin{figure}[!t]
    \centering
    \includegraphics[width=\linewidth]{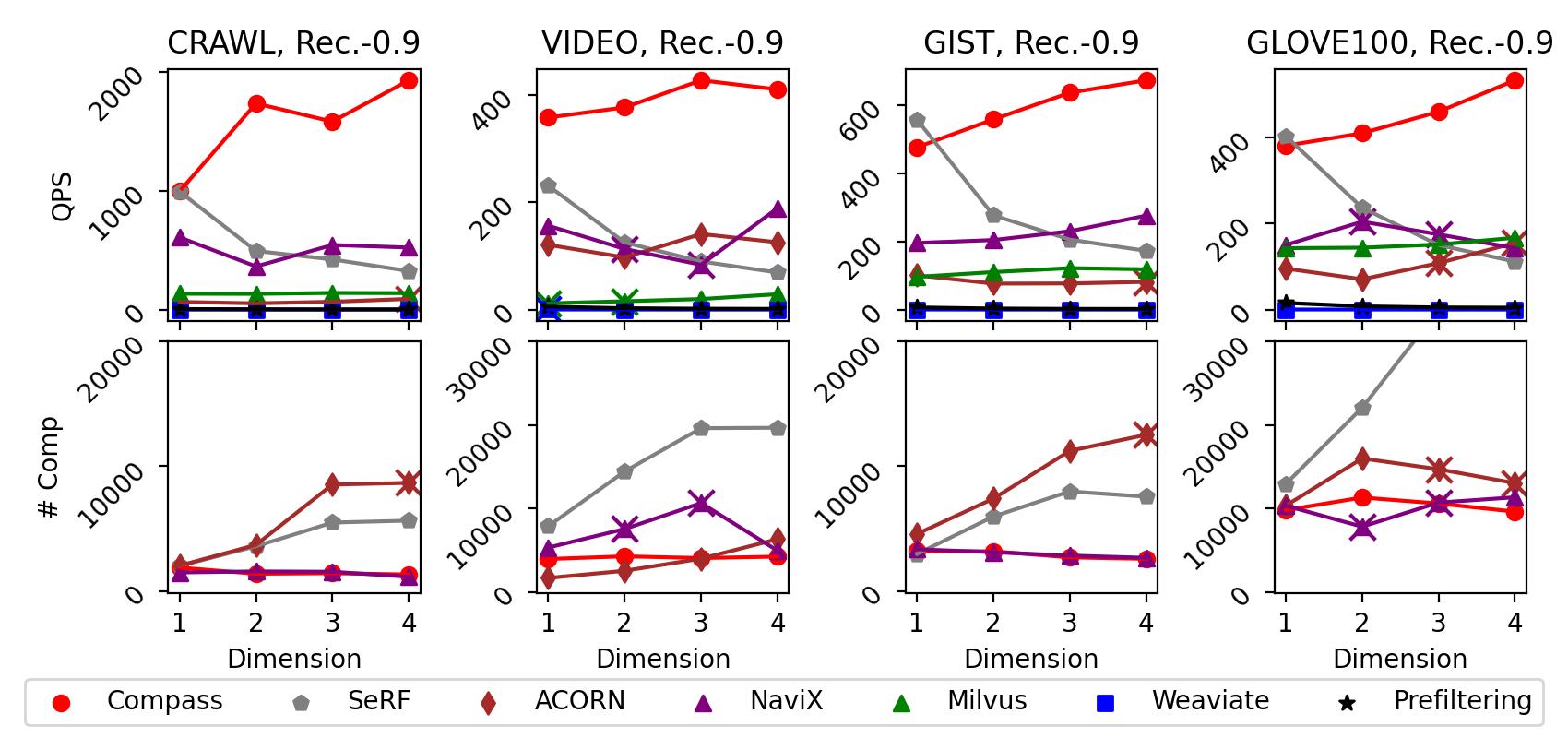}
    \caption{Disjunction Range Filtering. 0.9 Recall. }
    \label{fig:disjun-recall-0.9}
\end{figure}

\begin{figure}[!t]
    \centering
    \includegraphics[width=\linewidth]{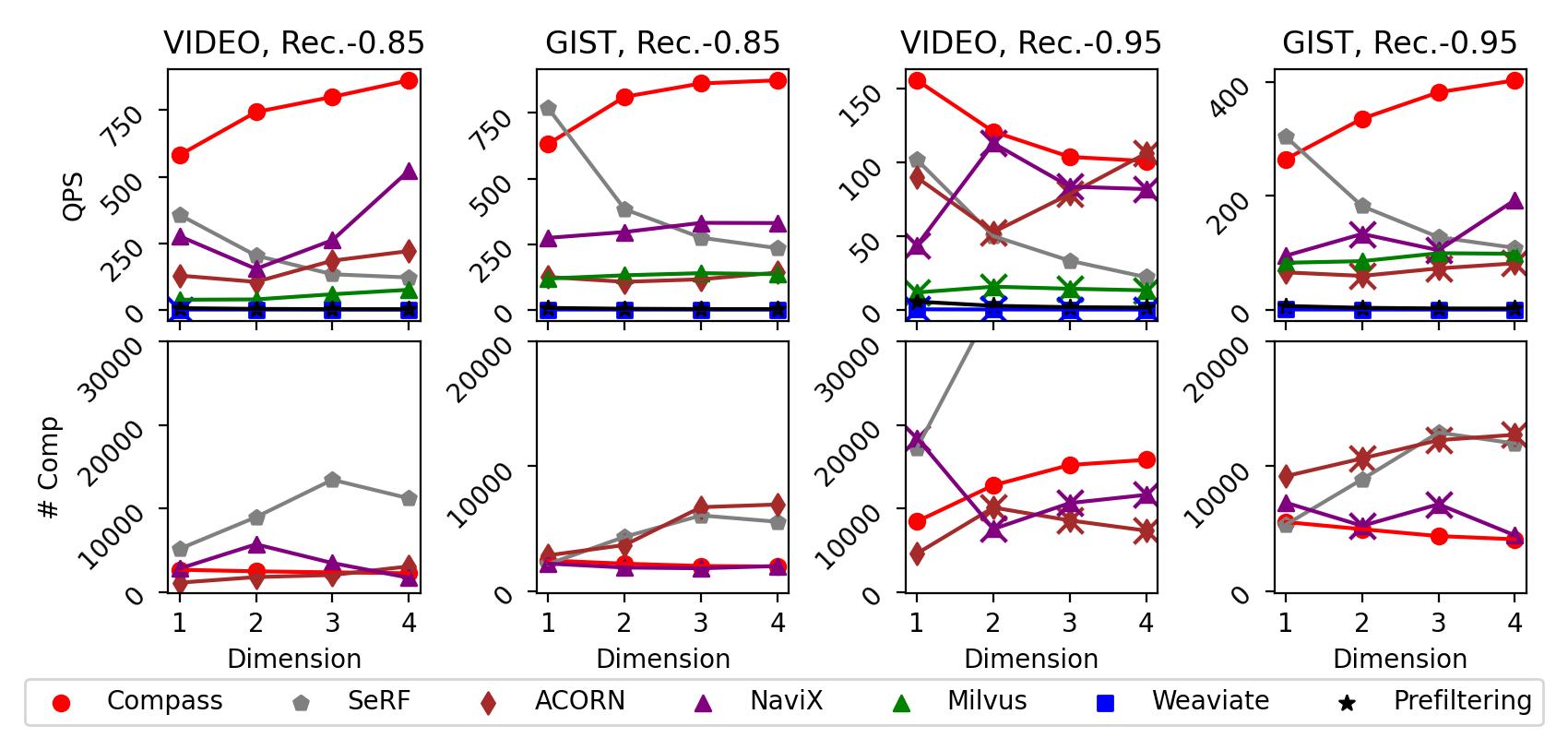}
    \caption{Disjunction Range Filtering. 0.85/0.95 Recall.}
    \label{fig:disjun-recall-0.85-0.95}
\end{figure}

As discussed in \cref{section:discussions},
while more selective conjunctions typically reduces the connectivity of the proximity graph (negatively impacting most baselines),
in Compass, more selective conjunctions actually decreases its reliance on the proximity graph.
Instead, it increasingly depends on the clustered B+-trees, which:
(1) are hardly affected by graph disconnections; and
(2) benefit from having fewer candidates.

\subsection{Disjunctions}

\Cref{fig:disjun-recall-0.9} presents the results using disjunctive predicates under recall 0.9 on all datasets while \Cref{fig:disjun-recall-0.85-0.95}
presents the results under recalls 0.85 and 0.95 for VIDEO and GIST only.
Since the default passrate of each attribute is 30\%,
the overall passrate for the disjunctive predicate increases
from 30\% with one attribute, to 51\% for two attributes, 66\% for three attributes
and 76\% for four attributes.

The results are consistent across all recall thresholds and align with the findings in conjunctions: Compass consistently outperforms NaviX and ACORN in QPS. 
ACORN is slower than NaviX in most cases due to ACORN's denser graph.
Furthermore, in comparison to conjunctions, Compass now significantly outperforms SeRF once beyond one-dimensional queries.
That is because SeRF requires one graph-index traversal per queried attribute,
after that, it unions the result and sorts the union to return $k$ results.
Milvus and Weaviate are not competitive, as they lack optimizations for filtered search. Prefiltering is entirely ineffective, as increasing the number of disjunction predicates significantly raises the number of vectors that need to be processed.

\subsection{QPS / \#Distance Computations vs Recall}\label{section:qps-recall}

\begin{figure}[!t]
    \centering
    \includegraphics[width=\linewidth]{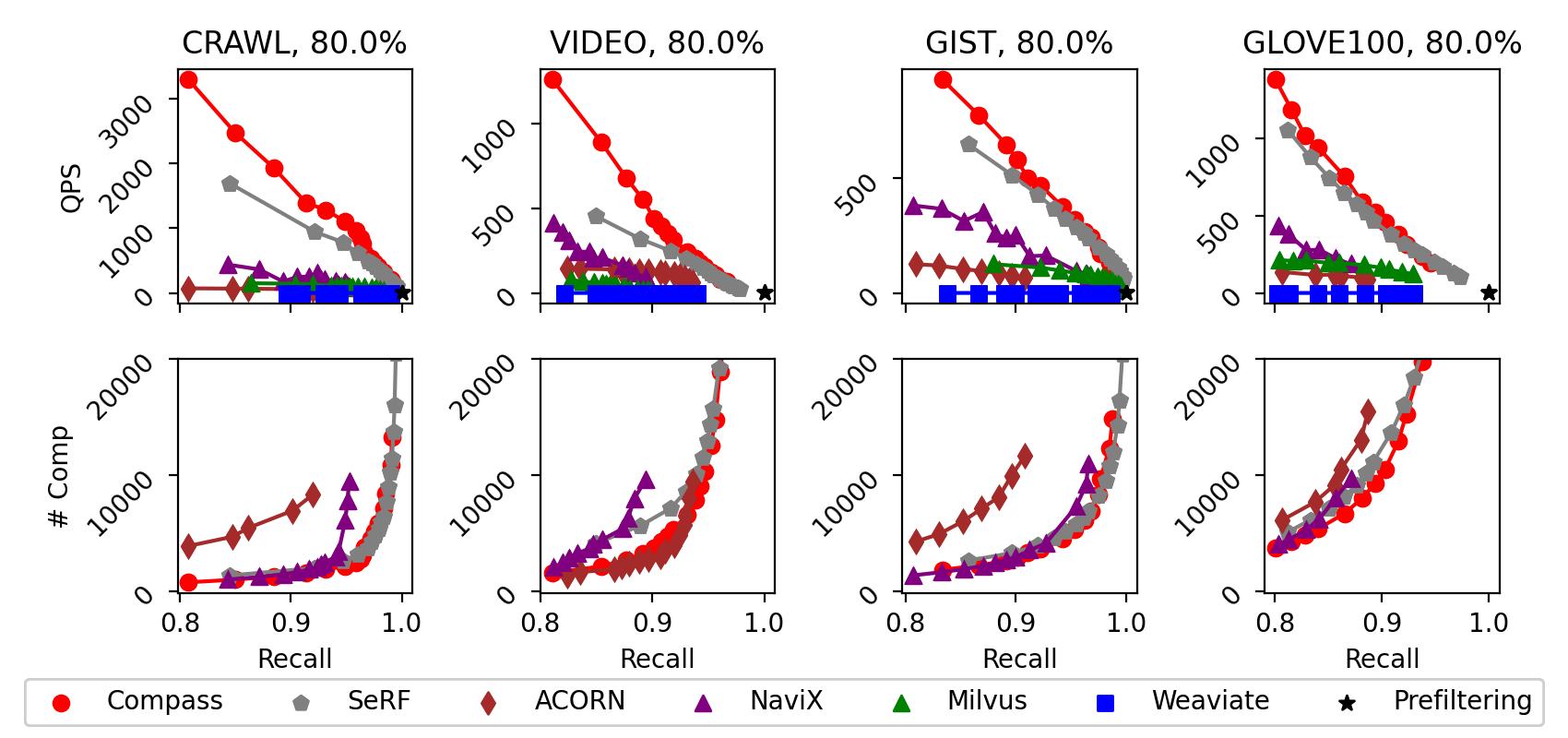}
    \caption{QPS and \#Computations vs Recall.  80\% passrate}
    \label{fig:recall-0.8}
\end{figure}

\begin{figure}[!t]
    \centering
    \includegraphics[width=\linewidth]{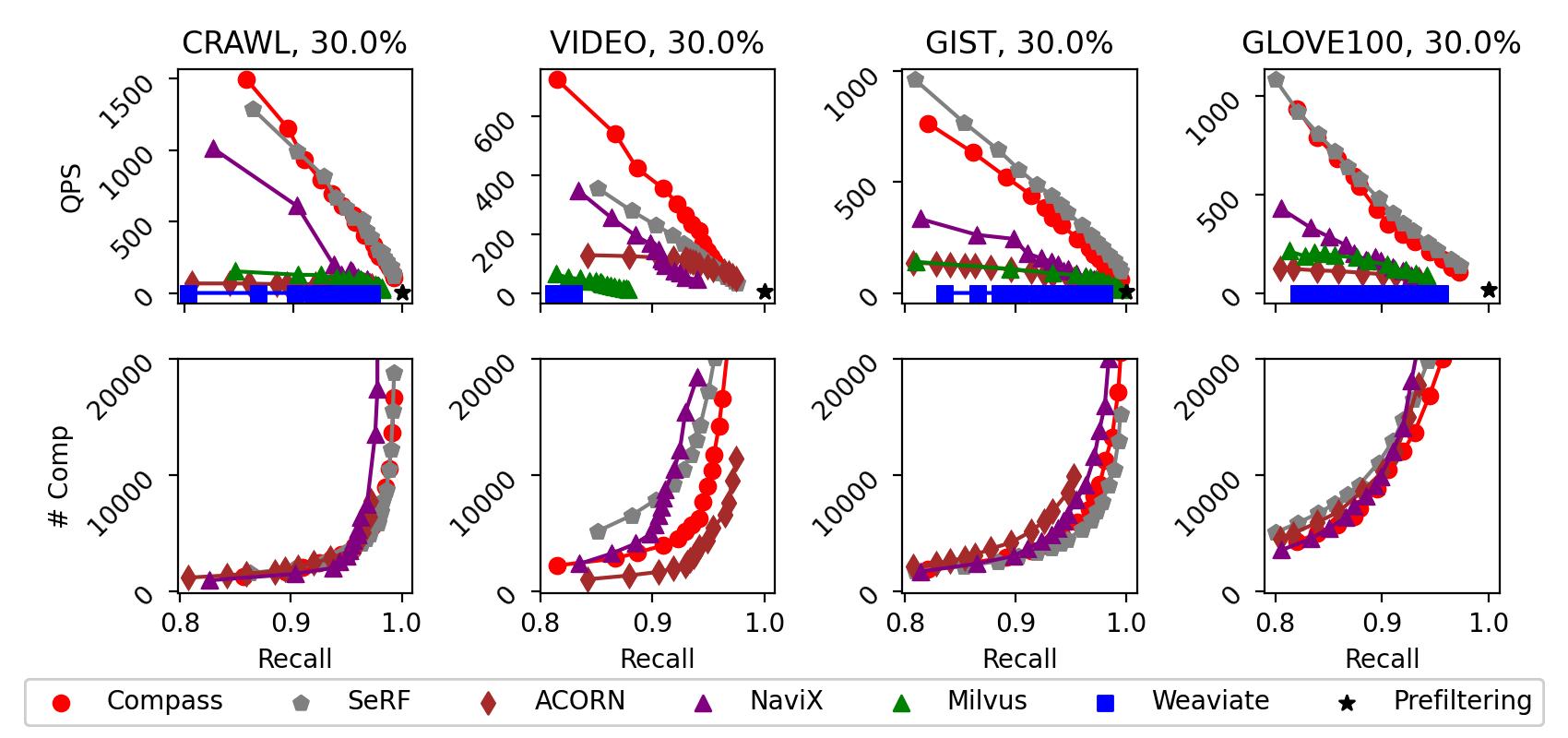}
    \caption{QPS and \#Computations vs Recall.  30\% passrate}
    \label{fig:recall-0.3}
\end{figure}

\begin{figure}[!t]
    \centering
    \includegraphics[width=\linewidth]{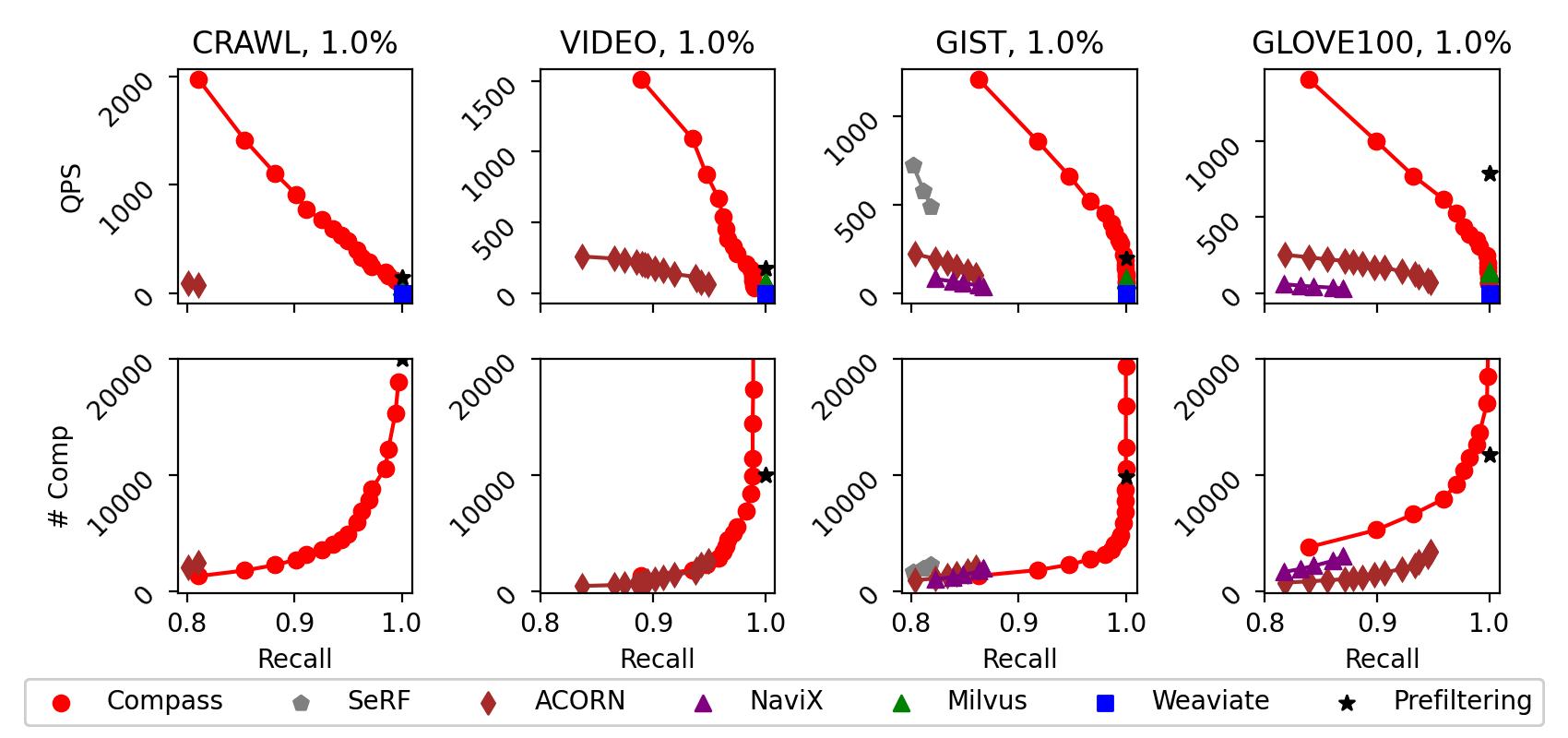}
    \caption{QPS and \#Computations vs Recall.  1\% passrate}
    \label{fig:recall-0.01}
\end{figure}

Figures \ref{fig:recall-0.8} to \ref{fig:recall-0.01} illustrate the QPS and the number of vector distance computations for all methods on a single attribute for recall ratios from 0.8 to 1.0 by varying \ef. Given that, for conjunction and disjunction, varying the number of attributes is equivalent to varying the selectivities, we present the results in three distinct selectivities for this experiment: an 80\% passrate (not selective), a 30\% passrate (default), and a 1\% passrate (selective).
From the figures,
we can see that only Compass can consistently return results with high recall across all three selectivities.

Under high passrate (\Cref{fig:recall-0.8}), NaviX and ACORN fail to stably return results with high recall ($\ge$ 0.9).
This is because under high passrate, they only compute vector distances for predicate-passing records,
a strategy that proves insufficient even when the graph connectivity is well-preserved.
In contrast, Compass also evaluates distances for non-passing neighbors, because a non-passing neighbor may itself have neighbors that are close to the query vector and pass the predicate,
which are potentially ignored by NaviX and ACORN.
Under low passrate (\Cref{fig:recall-0.01}),
NaviX, ACORN and SeRF 
have difficulty producing reasonable recall due to the graph disconnectivity. 
The proximity graph in Compass, on the other hand, can efficiently move to the other disconnected components due to the navigation from clustered B+-trees.
Milvus' and Weaviate's curve collapse near a single point because they choose the pre-filtering strategy at low passrate.

\subsection{Real Relational Attributes}\label{app:real}

In this section, we evaluate all methods using both real vectors and real relational attributes. Among the datasets considered, only the VIDEO dataset has real and standard relational attributes available: \texttt{'Watches'} and \texttt{'Likes'}.

\begin{figure}[!t]
    \centering
    \includegraphics[width=\linewidth]{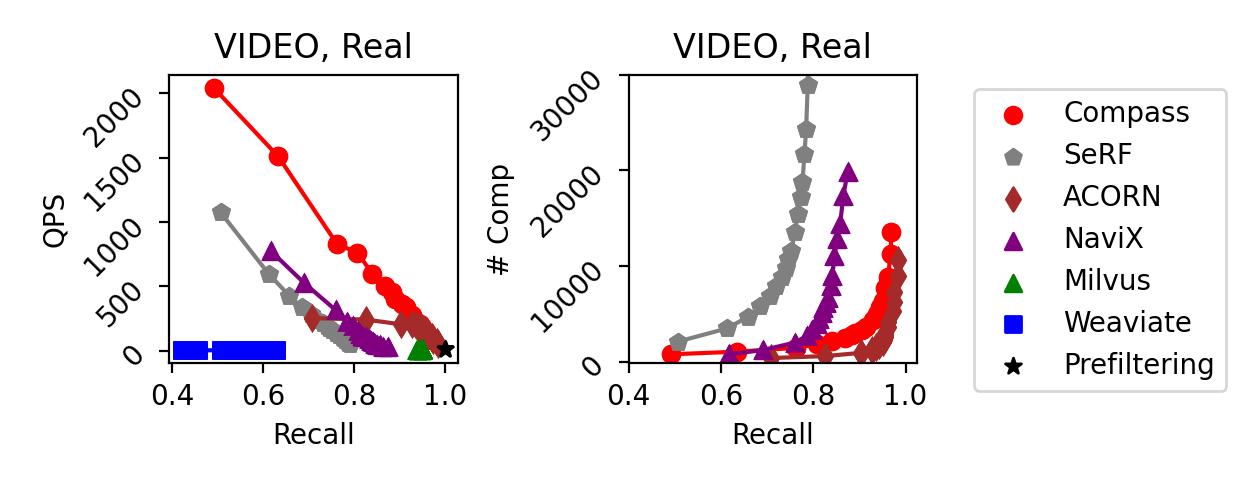}
    \caption{QPS and \#Computations vs Recall. Real Attribute.}
    \label{fig:real}
\end{figure}

We construct a query workload 
to retrieve videos similar to the query vector with more than 
\texttt{X} watches and \texttt{Y} likes, where 
\texttt{X} and \texttt{Y} are randomly sampled from the domains of their crawled values, resulting in an average passrate of around 13\%.

As shown in \Cref{fig:real}, Compass outperforms all the baseline methods in terms of the QPS-recall trade-off. While Compass performs a slightly larger number of distance computations than ACORN, it achieves better overall throughput. This is because ACORN only computes vector distance for predicate-passing records but consumes more time during the graph traversal and predicate evaluation due to its denser graph.

\subsection{More Results}
Due to limited space, we present additional experimental results in the appendix of our technical report~\cite{yeCompassGeneralFiltered2025}. These supplemental evaluations include:
\squishlist
\item \textbf{Various Attribute Distributions}: We evaluate across skewed, correlated and anti-correlated relational attributes.
\item \textbf{Diverse Filter Predicates}: We assess using one-sided range queries, equality matching (point queries), and negation filters.
\item \textbf{Scalability}: We evaluate on a larger memory-resident dataset DEEP10M of size 10 million, across various attribute distributions and diverse filter forms.
\squishend
Compass consistently outperforms baseline methods across all these evaluated workloads.
Furthermore, we conduct the following studies to validate the design integrity of Compass:
\squishlist
\item \textbf{Indexing Efficiency}: We provide a breakdown of Compass' construction time for graph, IVF, and B+-trees. The results indicate that Compass introduces negligible indexing overhead and scales well to the number of relational attributes.
\item \textbf{Ablation Study}: We verify Compass' structural cohesiveness by removing its core components: including the proximity graph, clustered B+-trees, progressive search mechanism, and cluster graph. Finally, we demonstrate that Compass remains robust and insensitive to its specific search parameters, namely $\deltaefs'$ and $\efi$.
\squishend

\section{Conclusions}\label{section:conclusion}

As a modular solution, \textsc{Compass} cohesively leverages existing indices with minimal intrusion into their underlying designs.
As a general-filter solution, its index construction is entirely predicate-agnostic, with all predicate-specific logic handled dynamically at search time. 
At the core of Compass is its adaptive search strategy,
which adjusts to the predicate passrate. 
When the passrate is high, the method operates as progressive search on the graph index. As the passrate transitions to a moderate level, it incorporates in-filtering techniques. When the passrate becomes low, the IVF component activates to help escape local minima. Throughout this process, Compass continuously balances two objectives: reducing vector distance and maintaining predicate satisfaction—a guided approach that inspired the name \emph{Compass}.

Compass can be extended to the disk-resident setting with only engineering changes:
(1) Using a disk-based B+tree and 
(2) Using a disk-resident graph index such as DiskANN~\cite{jayaramsubramanyaDiskANNFastAccurate2019}.
We note that the disk-resident setting itself is an interesting research topic \cite{wangStarlingOEfficientDiskResident2024, yinGorgeousRevisitingData2025}
when it comes to I/O optimization. This is complementary to our current focus on supporting filtered search, and we view that as an important future work.

\bibliographystyle{ACM-Reference-Format}
\bibliography{references/Citing, references/System-and-ML, references/System, references/eric, references/beautified}

\clearpage

\appendix

\section{Various Attribute Distributions}\label{app:attribute}

\begin{figure}[!t]
    \centering
    \includegraphics[width=\linewidth]{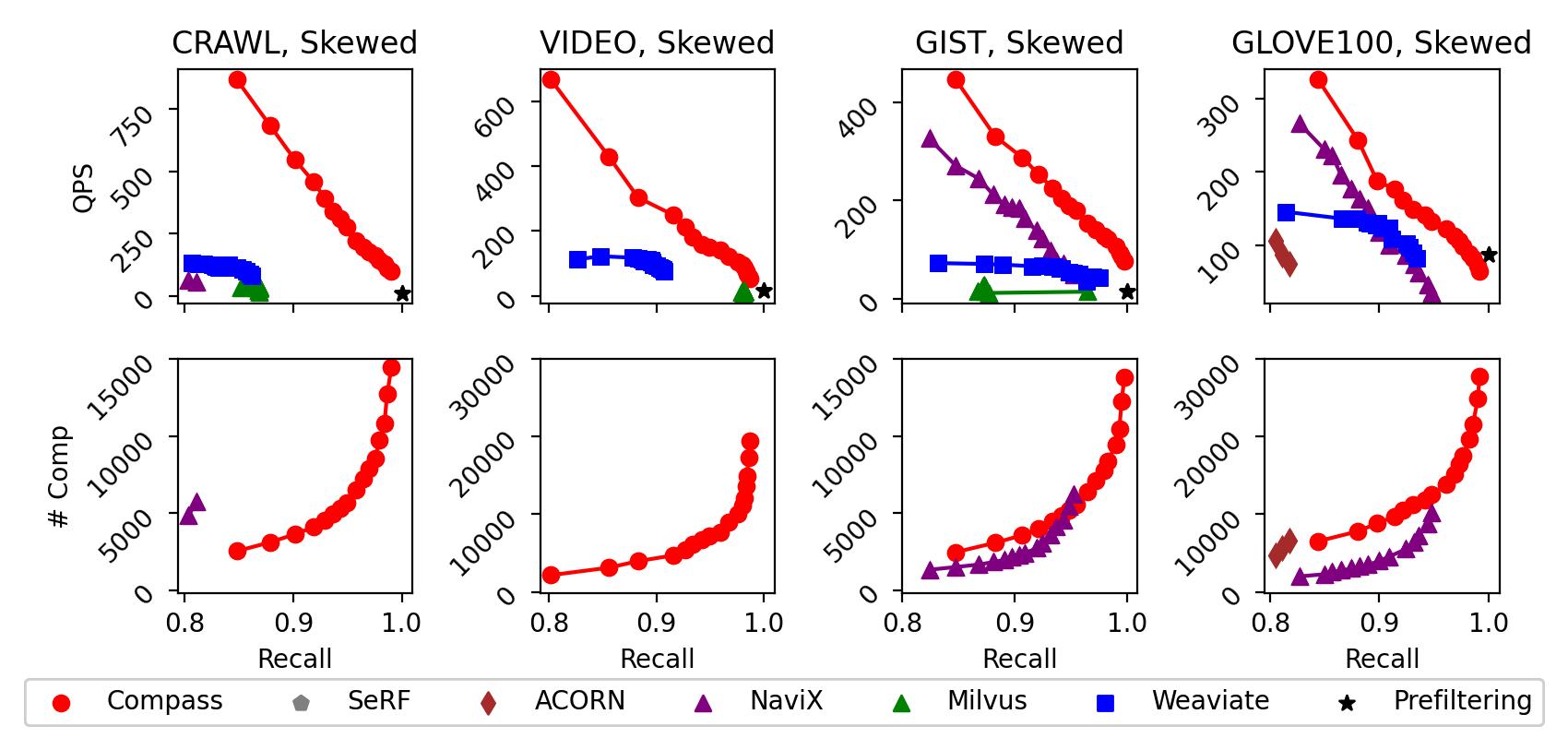}
    \caption{Skewed attribute.}
    \label{fig:skewed}
\end{figure}

\begin{figure}[!t]
    \centering
    \includegraphics[width=\linewidth]{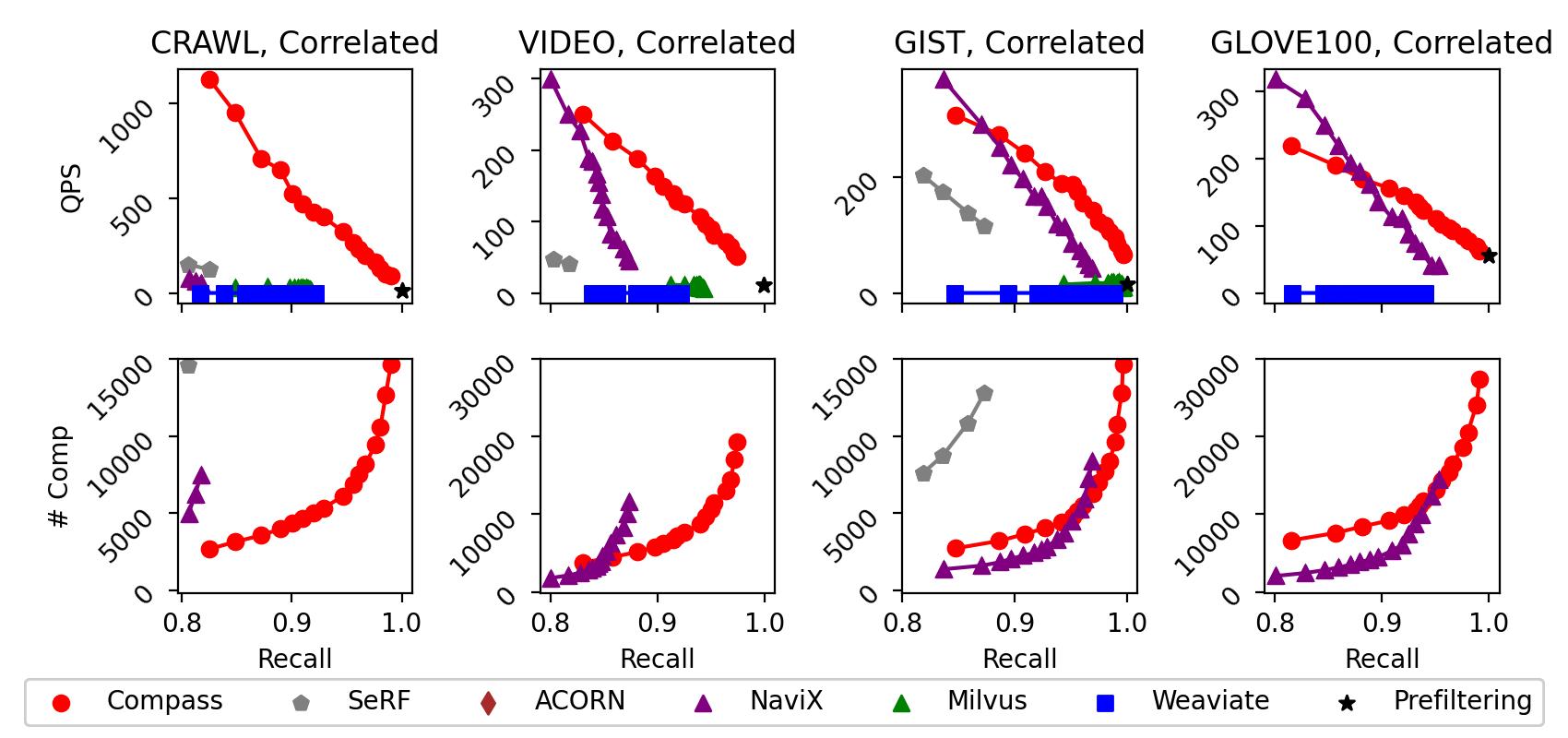}
    \caption{Correlated attributes.}
    \label{fig:correlated}
\end{figure}

\begin{figure}[!t]
    \centering
    \includegraphics[width=\linewidth]{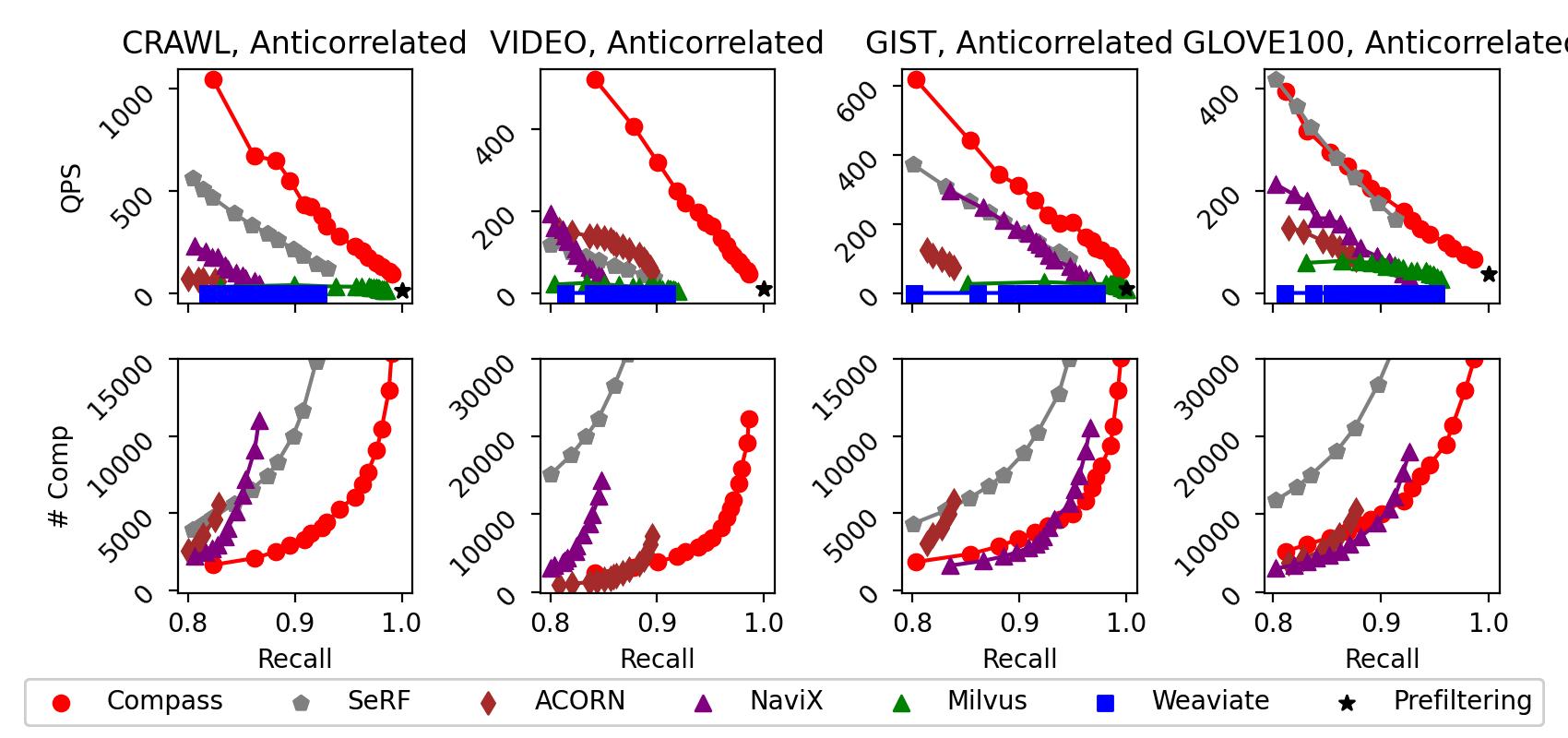}
    \caption{Anti-correlated attributes.}
    \label{fig:anticorrelated}
\end{figure}

In this experiment, we evaluate the query throughput and the number of distance computations for Compass, SeRF (with post-filtering), NaviX, ACORN, Milvus, Weaviate and Pre-filtering on skewed, correlated, anti-correlated attributes, respectively. 

Skewed attribute data is sampled from a Zipf distribution with parameter $\alpha$=2. Filtered search range queries are uniformly sampled within the interval $[0, 30]$, resulting in different passrates for different queries with an average of approximately 10\%.

For correlated attributes, data pairs are sampled from a two-dimensional Gaussian distribution with $\mu$=0, $\sigma$=10, and a correlation coefficient of 0.5.
Similarly, anti-correlated attribute pairs are sampled using a correlation coefficient of -0.5.

Conjunctive range queries are uniformly synthesized from the domain $[-20, 20] \times [-20, 20]$, yielding average passrates of approximately 10\% for the correlated workload and 15\% for the anti-correlated workload.

\Cref{fig:skewed}, \Cref{fig:correlated} and \Cref{fig:anticorrelated} present the results.
Only Compass can stably achieve high recall across all cases. Compass also excels in terms of the QPS-recall trade-off curve (upper right is better).

\begin{figure}[!t]
    \centering
    \includegraphics[width=\linewidth]{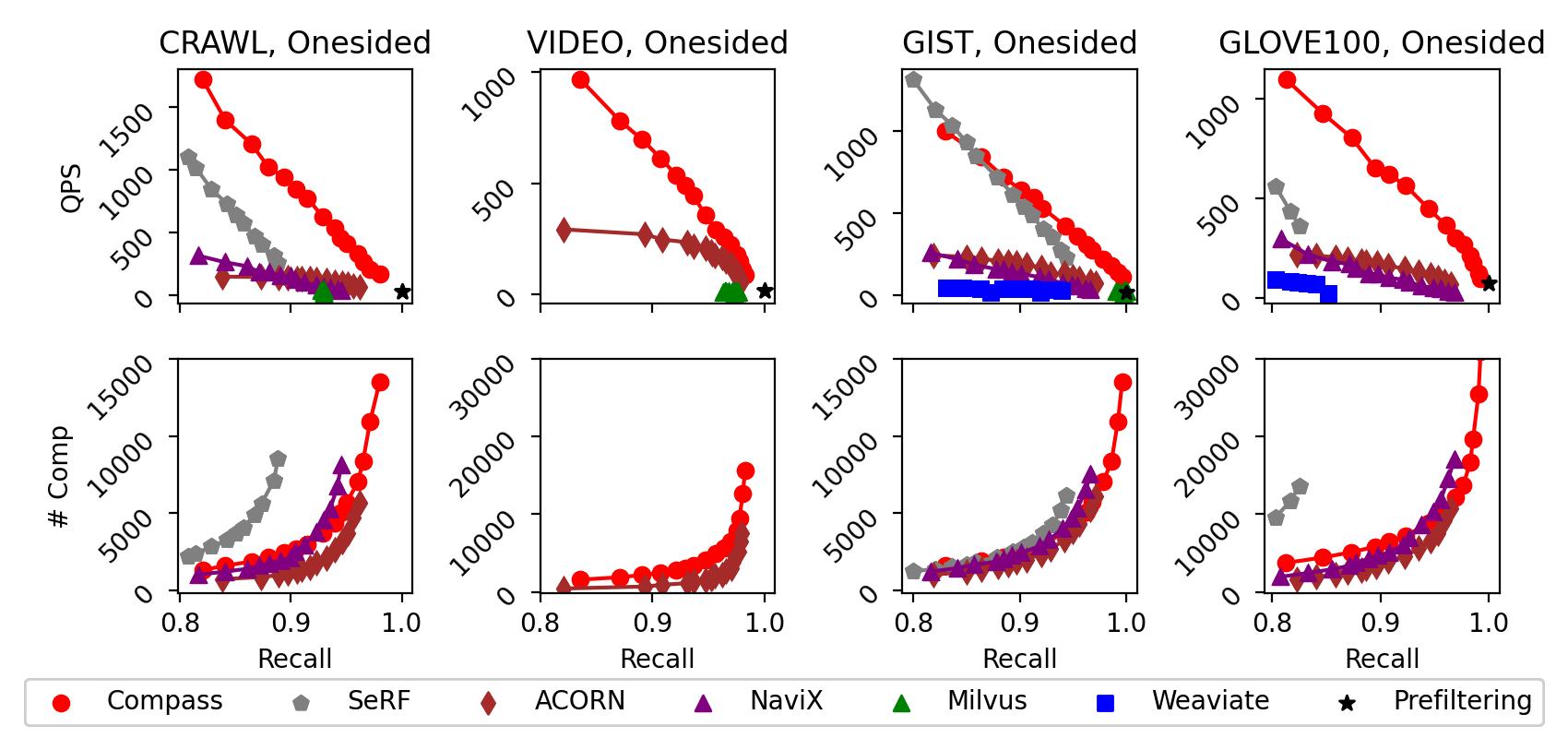}
    \caption{One-sided predicates.}
    \label{fig:onesided}
\end{figure}

\begin{figure}[!t]
    \centering
    \includegraphics[width=\linewidth]{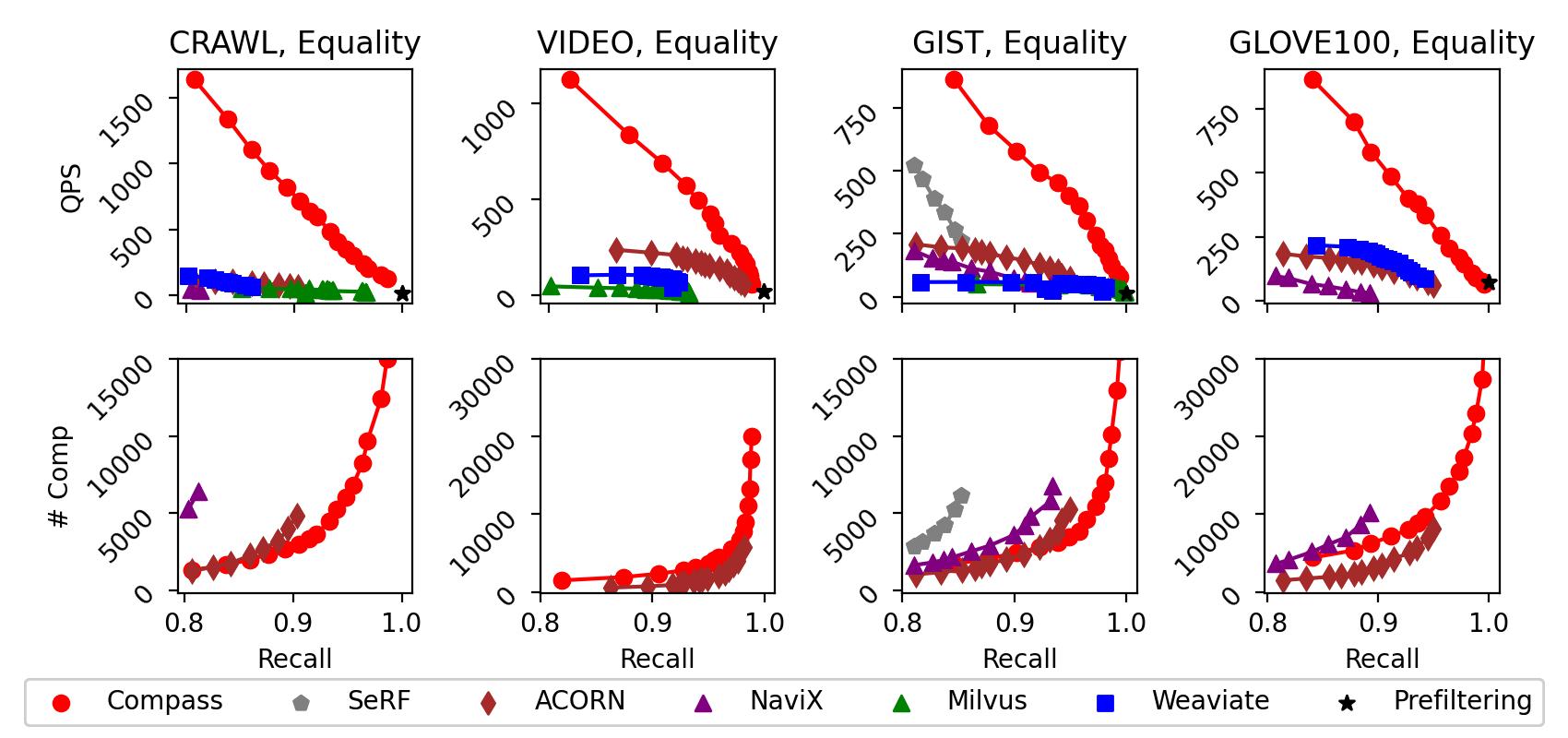}
    \caption{Equality predicates.}
    \label{fig:equality}
\end{figure}

\begin{figure}[!t]
    \centering
    \includegraphics[width=\linewidth]{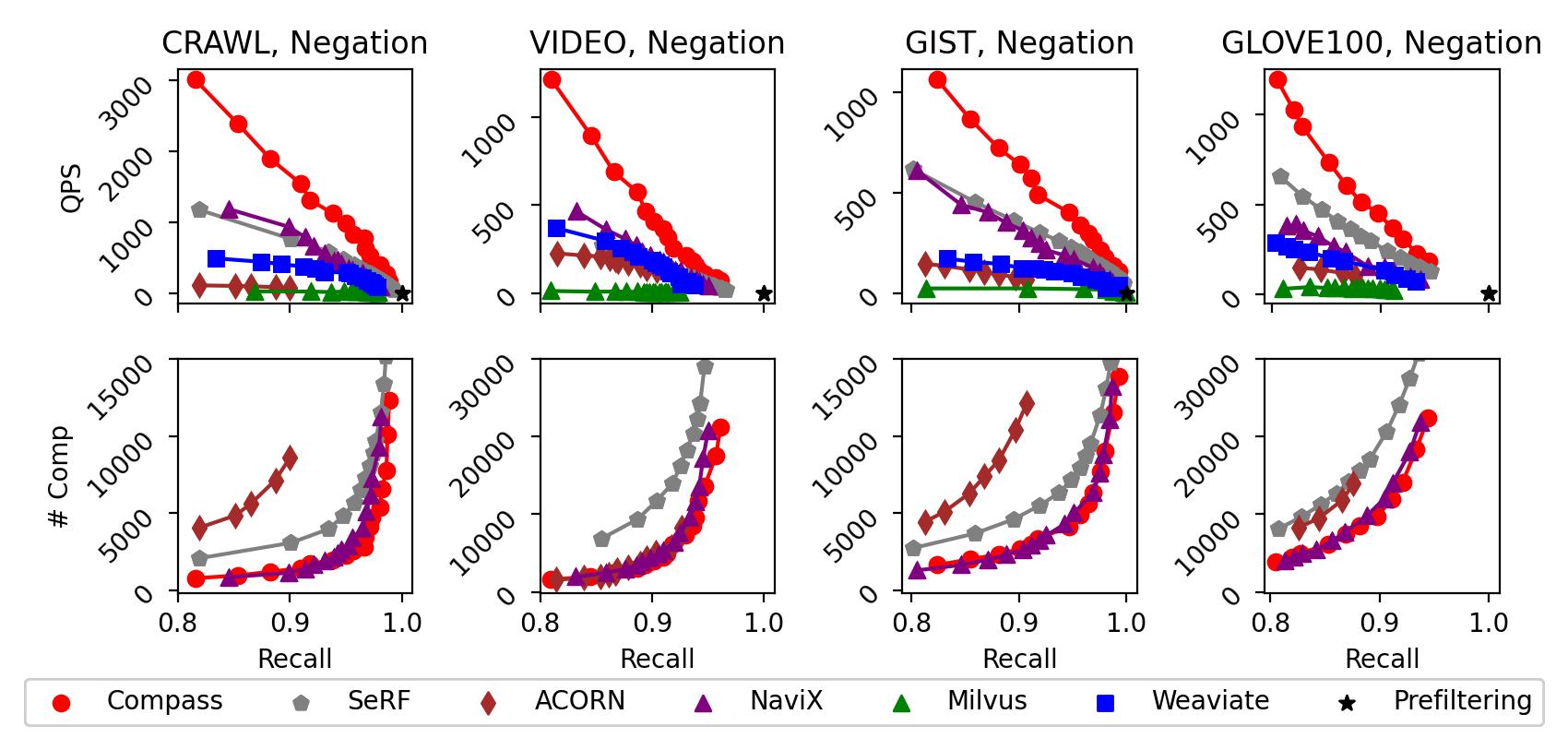}
    \caption{Negation predicates.}
    \label{fig:negation}
\end{figure}

\begin{table*}[t!]
\centering
\caption{Comparing the index size and construction time of \textbf{Compass} with baselines on DEEP10M.}

\resizebox{\textwidth}{!}{
\begin{tabular}{@{}ccccccccc@{}}
\toprule
\textbf{DEEP10M}     & \textbf{Compass}                                      & \textbf{SeRF}        & \textbf{NaviX$^\ddagger$}       & \textbf{ACORN}       & \textbf{Milvus$^\ddagger$}      & \textbf{Weaviate$^\ddagger$}    & \textbf{iRangeGraph$^\ddagger$} & \textbf{DSG}         \\
\multicolumn{1}{l}{} & \multicolumn{1}{l}{(\text{Graph$^\ddagger$ + IVF$^\ddagger$} + B+-trees)} & \multicolumn{1}{l}{} & \multicolumn{1}{l}{} & \multicolumn{1}{l}{} & \multicolumn{1}{l}{} & \multicolumn{1}{l}{} & \multicolumn{1}{l}{} & \multicolumn{1}{l}{} \\ \midrule
Size              & 2634+23+240*2=3137MiB                                 & 2.6*2=5.2GiB        & 1.1GiB              & 8.12GiB              & 2.6GiB$^*$               & N/A$^\dagger$                 & 22.8*2=45.6GiB       & 46.6*2=93.2GiB      \\ \midrule
Time              & 807+11871+1188=13866s                                  & 19739*2=39478s        & 1864s                 & 55026s                & 2226s                 & 2678s                 & 18886*2=37772s         & 103004*2=206008s      \\ 
\bottomrule
\end{tabular}
}
${}^*$: No function to provide index sizes.  Estimated according to https://milvus.io/tools/sizing.

${}^\dagger$: No function to provide index sizes. Not feasible to infer the index sizes because of its compact file storage.

${}^\ddagger$: Support multi-threading.  We use 32 threads.
\label{table:size-large}
\end{table*}

On skewed and correlated workload, though the average passrate is 10\%, the passrate distribution is skewed, towards the selective regions on which SeRF fail to achieve high recall due to its aggressive edge pruning, in accordance to the result in \Cref{fig:recall-0.01} in \Cref{section:conjunctions}.
This explains why SeRF fails to reach recall 0.8 on skewed workload.
Similarly, Milvus defaults to pre-filtering in these selective regions, causing its result points to concentrate rather than span the axis, even with the varying $\ef$.
However, for queries outside the selective region, Milvus' alternative simple strategy (e.g. in-filtering or post-filtering) proves non-effective, accounting for the performance gap between Milvus and other baselines.

Weaviate adopts an in-filtering strategy while ACORN adopts a simple two-hop traversal. They cannot stably achieve high recall across the distributions, and mostly under-perform NaviX, which has a more sophisticated traversal strategy. 
{Weaviate is extremely slow on correlated and anti-correlated distributions, potentially due to its row-oriented storage architecture, which manages multiple relational attributes as nested properties within individual records rather than in a columnar format.}

\section{Diverse Filter Forms}\label{app:diverse}

In this experiment, we evaluate the methods' generality across a diverse set of filter forms.
We utilize a single discrete relational attribute sampled from Zipf distribution with parameter $\alpha = 2$. We use skewed data for this experiment as it may present more realistic use cases compared to uniform data.

We constructed three query workloads: (i) one-sided inequality filters (\texttt{'> X'});
(ii) equality filters (\texttt{'= X'});
(iii) negation filters (\texttt{'!= X'}).
The query value \texttt{X} is sampled from the interval $[0, 30]$, $[0, 10]$, $[0, 10]$, resulting in average passrates of approximately 10\%, 10\%, and 90\%, respectively.

SeRF's design is constrained to single-interval range queries. To accommodate for SeRF, we map an equality filter (\texttt{'= X'}) to range query \texttt{'(X-1, X+1)'}; map a negation filter (\texttt{'!= X'}) to the union of two separate range queries (\texttt{'< X'} or \texttt{'> X'}).

\Cref{fig:onesided}, \Cref{fig:equality} and \Cref{fig:negation} present the results.
SeRF cannot consistently achieve high recall on one-sided inequality and equality workload.
While NaviX and ACORN occasionally match Compass in recall with comparable or lower number of distance computations, their traversal strategies necessitate predicate evaluation on a larger set of records, which degrades their overall QPS. Without special design, Milvus and Weaviate fail to achieve high recall or high QPS. Weaviate cannot achieve recall 0.8 on CRAWL and VIDEO for one-sided inequality.

Compass remains the only robust method across all the filter forms, and outperforms in the QPS-recall trade-off curve (upper right is better) at high recall levels.

\begin{table*}[t!]
\centering
\caption{Comparing \textbf{Compass'} construction time in seconds with baselines.}
\label{table:time}

\resizebox{\textwidth}{!}{
\begin{tabular}{@{}ccccccccc@{}}
\toprule
\textbf{Dataset}     & \textbf{Compass}                                       & \textbf{SeRF}        & \textbf{NaviX$^\ddagger$}      & \textbf{ACORN}       & \textbf{Milvus$^\ddagger$}     & \textbf{Weaviate$^\ddagger$}   & \textbf{iRangeGraph$^\ddagger$} & \textbf{DSG}         \\
\multicolumn{1}{l}{} & \multicolumn{1}{l}{(Graph$^\ddagger$ + IVF$^\ddagger$ + B+-trees)} & \multicolumn{1}{l}{} & \multicolumn{1}{l}{} & \multicolumn{1}{l}{} & \multicolumn{1}{l}{} & \multicolumn{1}{l}{} & \multicolumn{1}{l}{}  & \multicolumn{1}{l}{} \\ \midrule
CRAWL                & 207+673+70=950                                       & 3122*4=12488         & 312                  & 4837                 & 410                  & 632                  & 4358*4=17432          & 10645*4=42580        \\
GIST                 & 192+581+58=831                                       & 2863*4=11452         & 319                  & 5693                 & 691                  & 566                  & 3211*4=12844          & 9531*4=38124         \\
VIDEO                & 430+1234+124=1788                                     & 4073*4=16292         & 309                  & 8683                 & 733                  & 592                  & 4488*4=17952          & 6028*4=24112         \\
GLOVE100             & 79+592+59=730                                        & 1187*4=4748          & 185                  & 3581                 & 238                  & 314                  & 2179*4=8716           & 3659*4=14636         \\ \midrule
\end{tabular}
}

${}^\ddagger$ Support multi-threading.  We use 32 threads.
\end{table*}

\section{Larger Dataset}\label{app:larger-data}

\begin{figure}[!t]
    \centering
    \includegraphics[width=\linewidth]{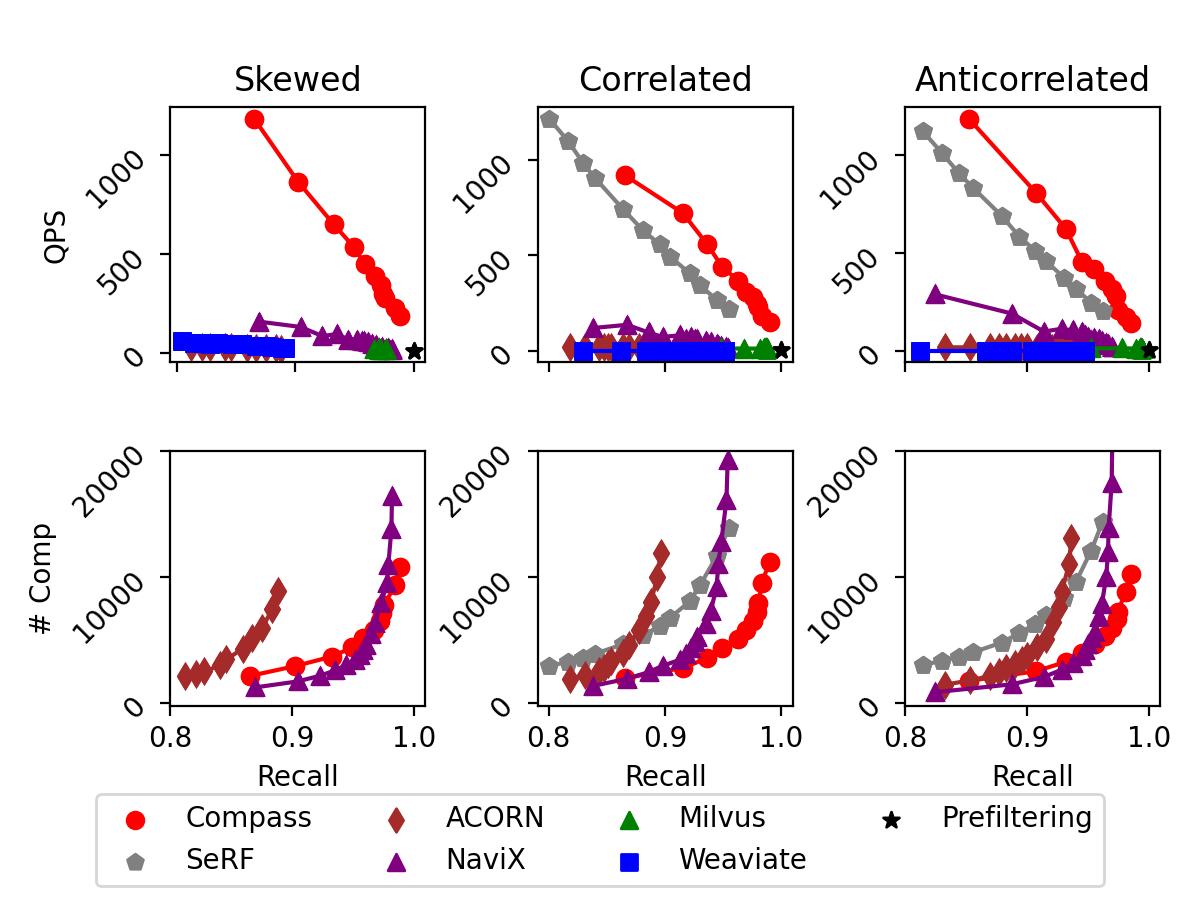}
    \caption{Large dataset: different attribute distributions.}
    \label{fig:deep10m-distrib}
\end{figure}

\begin{figure}[!t]
    \centering
    \includegraphics[width=\linewidth]{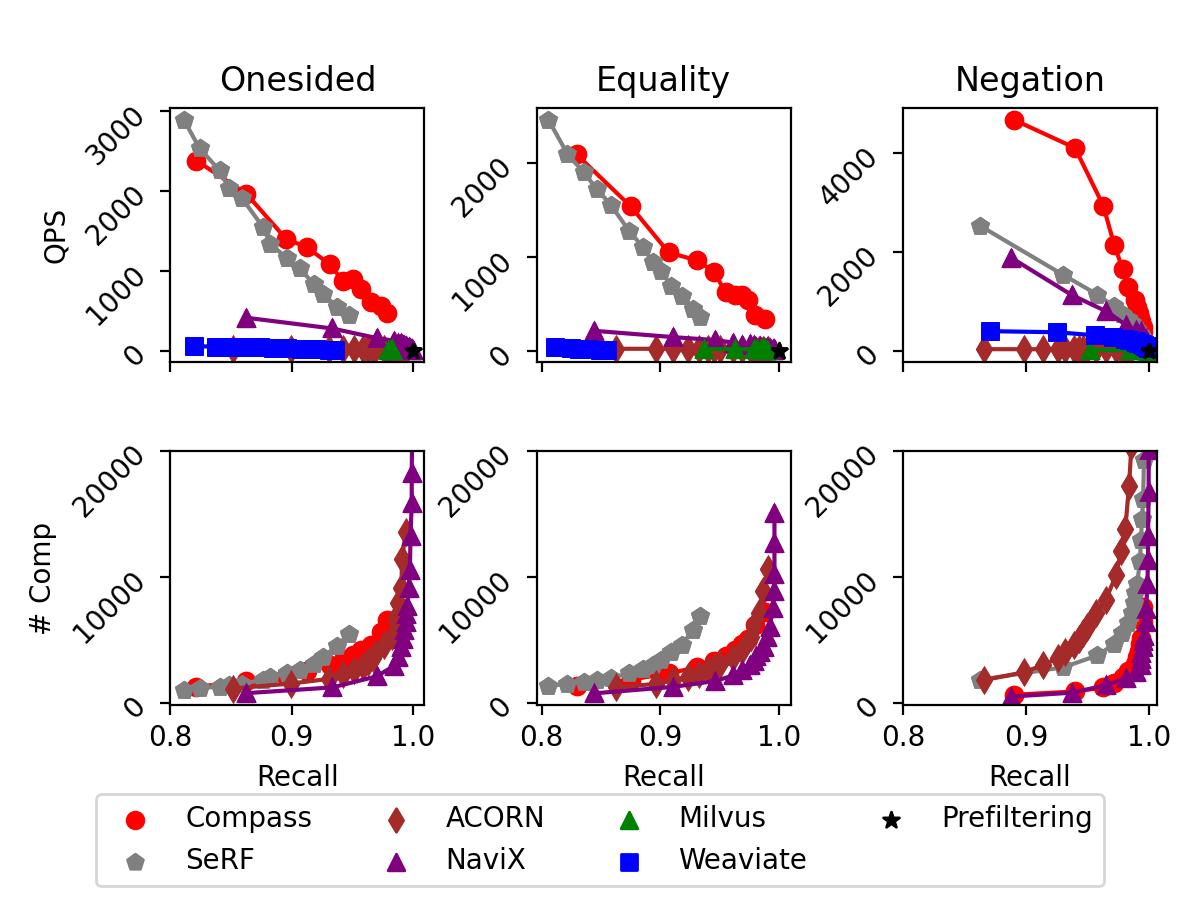}
    \caption{Large dataset: different filter forms.}
    \label{fig:deep10m-workload}
\end{figure}

In this experiment, we evaluate the scalability using the larger but still memory-resident dataset DEEP10M\footnote{https://research.yandex.com/blog/benchmarks-for-billion-scale-similarity-search} (containing 10,000,000 96-dimensional image embedding vectors), as commonly used for scalability analysis~\cite{zuoSeRFSegmentGraph2024, xuIRangeGraphImprovisingRangededicated2024, malkovEfficientRobustApproximate2020, fuHighDimensionalSimilarity2022}. 
Due to the unavailability of the source content, relational attributes and queries were synthesized following Appendix~\ref{app:attribute} and Appendix~\ref{app:diverse}.

For this evaluation, we set \textit{M}=32 for the base HNSW in all the methods (\textit{M}=64 for SeRF correspondingly). For Compass, we set \textit{nlist}=50,000 for IVF and \textit{M'}=8 for cluster graph.
All other index construction and search parameters remain the same as in our main evaluation.

\Cref{fig:deep10m-distrib} presents the QPS and number of distance computations relative to recall for skewed and non-independent attributes; \Cref{fig:deep10m-workload} presents those for diverse filter forms.
Compass still maintains high performance in QPS (upper right is better) and robustness across all the distributions and workloads on the large-scale DEEP10M dataset. We note that SeRF still cannot achieve recall 0.8 on skewed distribution. 
In one-sided inequality and equality workload that involve single-attribute predicates—conditions typically favorable to SeRF—Compass achieves performance comparable to, or surpassing, that of SeRF. Other methods all under-perform Compass.

Meanwhile, Compass requires much smaller index storage and shorter construction time (\Cref{table:size-large}) than SeRF, ACORN, iRangeGraph and DSG.

\section{Construction Time}\label{app:build-time}

In this section, we report the index construction time for all the methods, as illustrated in \Cref{table:time}.
Note that the costs reported here should be interpreted with caution
because some implementations use multi-threading but some do not.

For Compass construction, we utilize the HNSW from hnswlib~\cite{malkovEfficientRobustApproximate2020} (multi-threaded); the K-means algorithm from Faiss library~\cite{douzeFaissLibrary2024} (multi-threaded); and the B+-tree from~\cite{frozencaFrozencaBTree2026} (single-threaded).

In general, indices designed for filtered search—such as ACORN, SeRF, iRangeGraph, DSG, and Compass—incur higher construction costs compared to generic indices like NaviX, Milvus, and Weaviate. 
This increased construction latency is a justified trade-off to support efficient filtered search at runtime.

Among these filtered-search-optimized methods, Compass achieves the lowest construction time. While a significant portion of this time is dedicated to the IVF (Inverted File Index) clustering process - specifically K-means - this is purely an offline process that enables Compass to deliver robust and efficient filtered search performance at runtime. In practice, hierachical way of K-means construction~\cite{chenSPANNHighlyefficientBillionscale2021} can be adopted or Faiss IVF can benefit from GPU acceleration, which can further reduce the construction time. Besides, only a small portion of Compass' construction time (B+-trees) is dedicated to relational data, demonstrating its scalability to the number of relational attributes.

\section{Ablation Study}\label{app:ablation}

\begin{figure}[!t]
    \centering
    \includegraphics[width=\linewidth]{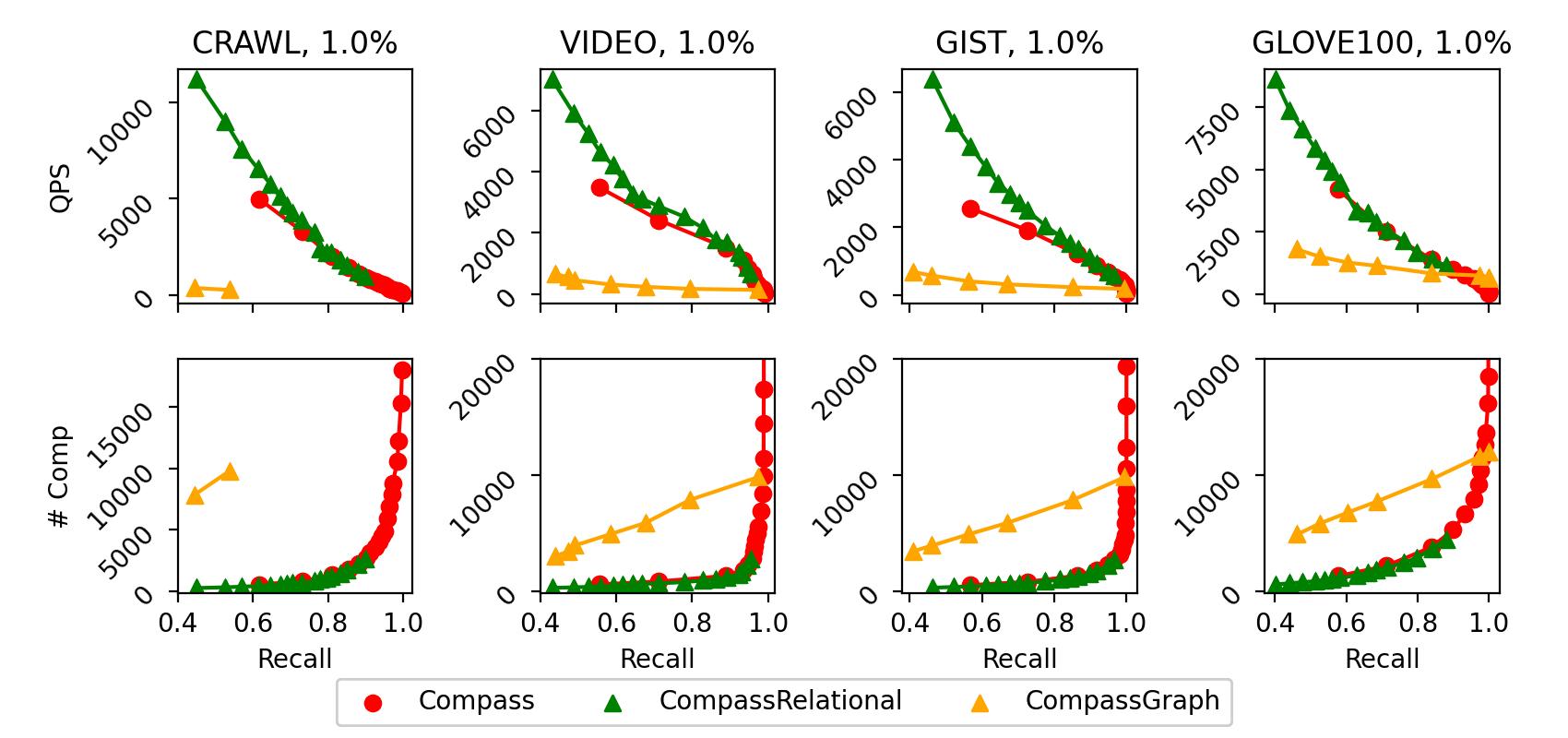}
    \caption{Ablation: component ablation (i)}
    \label{fig:ablate-0.01}
\end{figure}

\begin{figure}[!t]
    \centering
    \includegraphics[width=\linewidth]{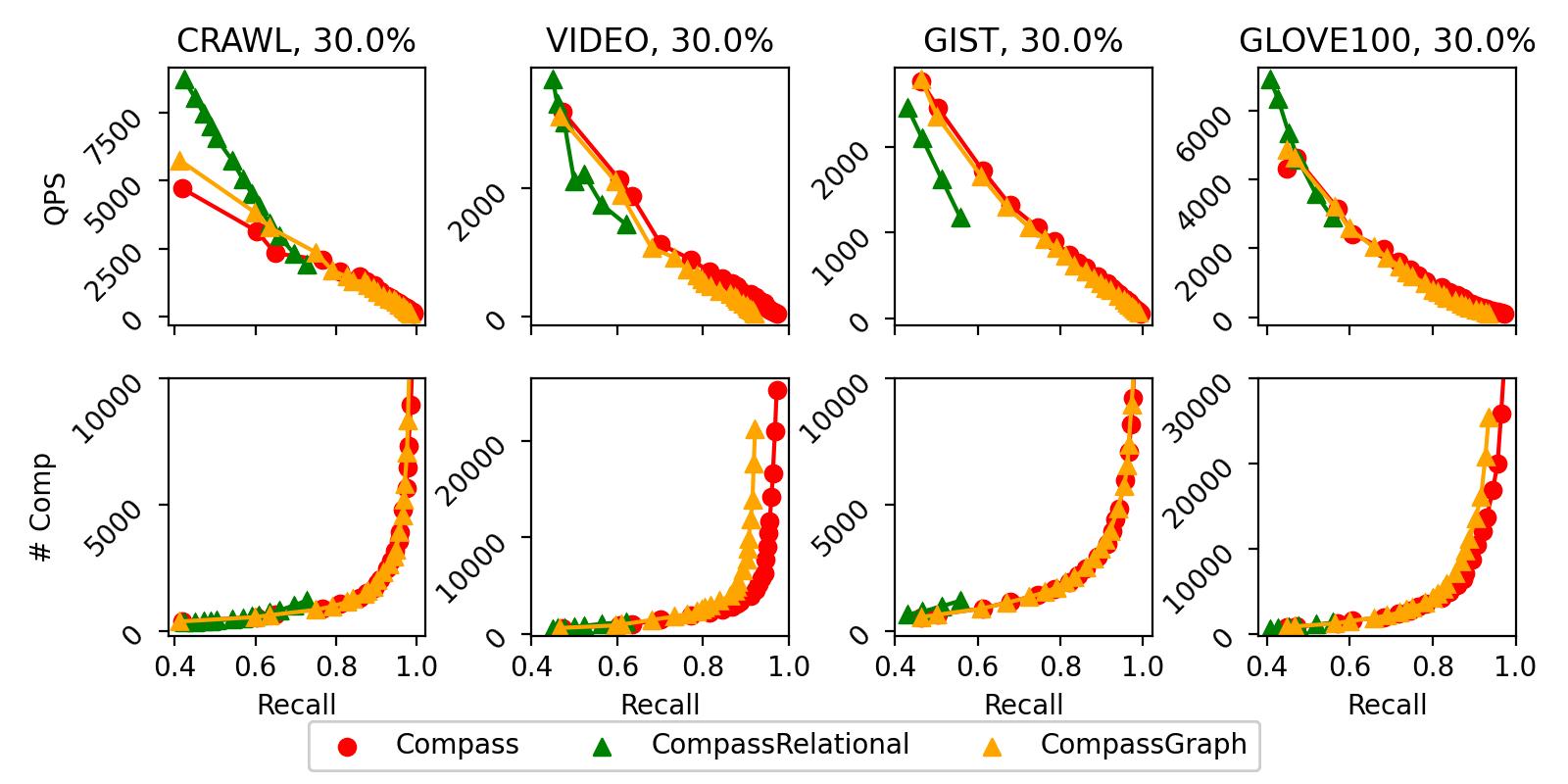}
    \caption{Ablation: component ablation (ii)}
    \label{fig:ablate-0.3}
\end{figure}

\stitle{Component Ablation} In this ablation study, we evaluate the contribution of individual components by examining two variants of Compass: \textbf{CompassRelational} and \textbf{CompassGraph}.
The CompassRelational variant is obtained by removing the proximity graph component, relying solely on the clustered B+-trees to progressively fetch candidates from clusters close to the query vector.
The CompassGraph variant is obtained by setting the number of clusters \textit{nlist}=1, effectively reducing the architecture to a single global B+-tree constructed on the attribute values of the entire dataset, thereby eliminating the cluster-based navigation.

\Cref{fig:ablate-0.01} and \Cref{fig:ablate-0.3} illustrate the QPS and the number of vector distance computations for recall levels from 0.4 to 1.0 by varying \ef from 10 to 1000  for this study;
their passrates are the selective passrate, 1\%, and the default passrate, $30\%$, respectively. All other index construction and search parameters remain the same as in our main evaluation.

In \Cref{fig:ablate-0.01}, CompassGraph cannot return result with high throughput. This is because the single global B+-tree provides filtering capability but lacks the geometric locality guarantees offered by the clustering mechanism. While it can still iterate records satisfying the relational predicate, it cannot prioritize candidates based on vector proximity. Without this navigation, the search is unable to efficiently overcome the induced graph's disconnectivity.

In \Cref{fig:ablate-0.3}, CompassRelational cannot return results with high recall across any of the four datasets because it lacks the proximity graph to efficiently approach the query vector.

This ablation study demonstrates that the proximity graph and the clustered B+-trees are integral and complementary components of Compass.

\begin{figure}[!t]
    \centering
    \includegraphics[width=\linewidth]{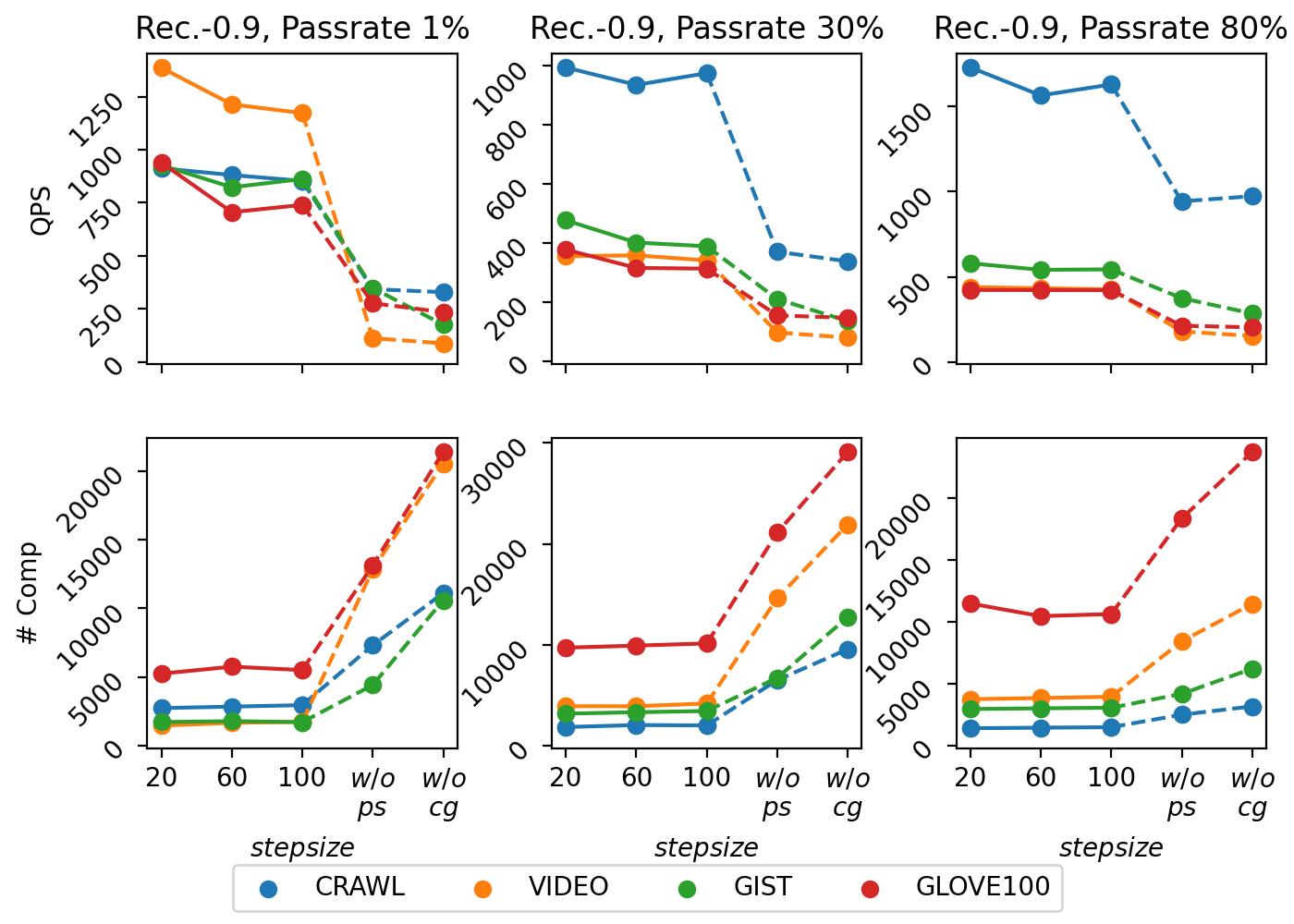}
    \caption{Ablation: progressive search.}
    \label{fig:ablation-progressive-search}
\end{figure}

\begin{figure}[!t]
    \centering
    \includegraphics[width=\linewidth]{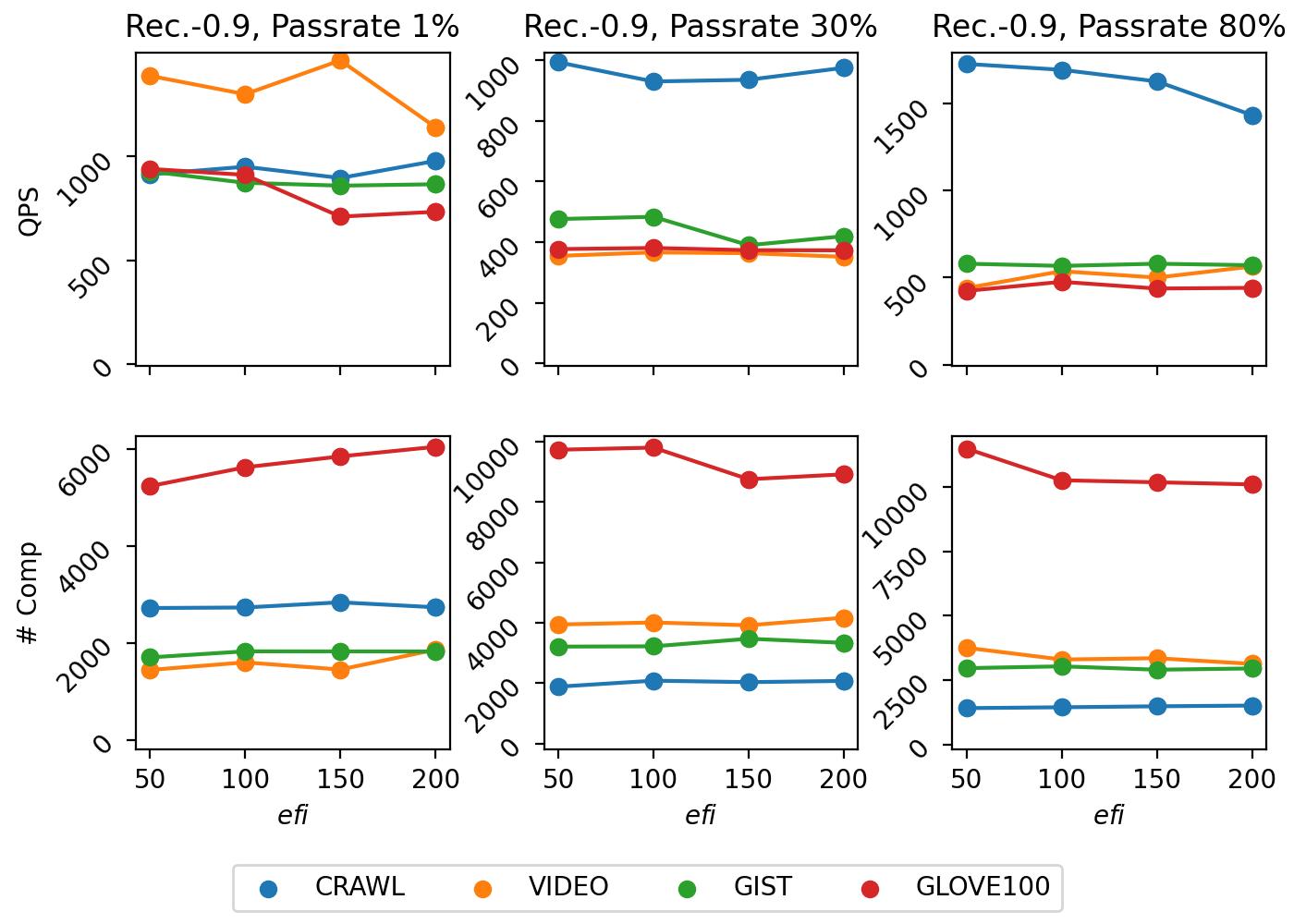}
    \caption{Ablation: $\efi$.}
    \label{fig:ablation-nrel}
\end{figure}

\stitle{Parameter Sensitivity}
In this ablation study, we discuss the rationale behind the setting of two Compass-specific search parameters, i.e. $\deltaefs'$ and $\efi$, and show that Compass is largely insensitive to their variations.
Additionally, we validate the necessity of progressive search and cluster graph in our design.

To recall, $\deltaefs$ is the increment step to enlarge $\efs$ as well as the number of candidates to be returned from the graph in each iteration.
For proximity graph $\graph$, we fix $\deltaefs$=$k$, aiming to directly retrieve the approximate top-$k$ if $\graph$ is sufficiently connected.
For cluster graph $\cg$, if $\deltaefs'$ is too small, the iteration would end without returning enough closer clusters; if too large, unnecessary computations would be wasted.

However, we note that the setting of $\deltaefs'$ causes only slight performance variation. As shown in \Cref{fig:ablation-progressive-search}, varying $\deltaefs'$ for the cluster graph from 20 to 100 results in stable QPS and distance computation metrics across datasets and across passrates of 1\% (selective), 30\% (default), and 80\% (non-selective), for a target recall of 0.9.
Based on the these findings, we set $\deltaefs'$=20 for the cluster graph $\cg$ during evaluation, identifying it as a balance point between the result quality and search efficiency.

We further evaluate the structural design by ablating the progressive search (``\textit{w/o ps}''), which replaces the progressive process with a fixed-size retrieval of close clusters from the cluster graph, and ablating the cluster graph (``\textit{w/o cg}''), which finds the fixed-size close clusters by brute-force comparison.
Both ablations cause a large increase in vector distance computations and a corresponding drop in QPS, confirming that the cluster graph and progressive search are critical to the system's efficiency.

Finally, the parameter $\efi$ is designed to align the computational cost of a single proximity graph traversal iteration with that of a single clustered B+-tree retrieval iteration.
As shown in \Cref{fig:ablation-nrel}, variations in $\efi$ cause little performance fluctuations on all the datasets across passrates 1\%, 30\%, 80\%, for a target recall of 0.9.
We selected the best setting of $\efi$=50 on CRAWL, VIDEO, GIST, and $\efi$=100 on GLOVE100 during evaluation.

\end{document}